\documentclass[conference]{IEEEtran}
\IEEEoverridecommandlockouts

\usepackage{academicons}
\usepackage{cite}
\usepackage{amsmath,amssymb,amsfonts}
\usepackage{algorithmic}
\usepackage{graphicx}
\usepackage{textcomp}
\usepackage{xcolor}
\usepackage{stfloats}
\usepackage{multirow}

\def\BibTeX{{\rm B\kern-.05em{\sc i\kern-.025em b}\kern-.08em
    T\kern-.1667em\lower.7ex\hbox{E}\kern-.125emX}}
\begin{document}

\title{A Novel Power-optimized CMOS sEMG Device with Ultra Low-noise integrated with ConvNet (VGG16) for Biomedical Applications}

\author{\IEEEauthorblockN{Ahmed Ayman}
    \IEEEauthorblockA{\textit{El-Sadat STEM School} \\
        Menoufia, Egypt}\\
    \and
    \IEEEauthorblockN{Mohamed Sabry}
    \IEEEauthorblockA{\textit{El-Sadat STEM School} \\
        Menoufia, Egypt}
}

\maketitle

\begin{abstract}
    The needle bio-potential sensors for measuring muscle and brain activity need invasive surgical targeted muscle reinnervation (TMR) and a demanding process to maintain, 
    but surface bio-potential sensors lack clear bio-signal reading (Signal-Interference). In this research, a novel power-optimized complementary metal-oxide-semiconductor 
    (CMOS) Surface Electromyography (sEMG) is developed to improve the efficiency and quality of captured bio-signal for biomedical application: The early diagnosis of neurological 
    disorders (Dystonia) and a novel compatible mind-controlled prosthetic leg with human daily activities. A novel sEMG composed of CMOS Op-Amp based PIC16F877A 8-bit CMOS Flash-based 
    Microcontroller is utilized to minimize power consumption and data processing time. sEMG Circuit is implemented with developed analog filter along with infinite impulse response (IIR) 
    digital filter via Fast Fourier Transform (FFT), Z-transform, and difference equations. The analysis shows a significant improvement of 169.2\% noise-reduction in recorded EMG signal 
    using developed digital filter compared to analog one according to numerical root mean square error (RMSE). Moreover, digital IIR was tested in two stages: algorithmic and real-world. 
    As a result, IIR's algorithmic (MATLAB) and real-world RMSEs were 0.03616 and 0.05224, respectively. A notable advancement of 20.8\% in data processing duration in sEMG signal analysis. 
    Optimizing VGG, AlexNet, and ResNet ConvNet as trained and tested on 15 subjects' public EEG (62 electrodes) and observed EMG data. The results indicate that VGG16-1D is 98.43\% higher. 
    During real testing, the accuracy was 95.8 ± 4.6\% for 16 subjects (6 Amputees-10 Dystonia). This study demonstrates the potential for sEMG, paving the way for biomedical applications.
\end{abstract}


\section{Introduction}
Electromyography (EMG) is a valuable diagnostic technique used in modern healthcare settings to assess muscle electrical activity. However, Surface EMG (sEMG) technology faces various obstacles, includes: 
The complex interpretation of sEMG signals, which can be challenging due to factors such as noise and interference. Additionally, 
sEMG patterns can vary significantly among individuals and muscles, making it challenging to establish standardized norms for interpretation. 
Regarding the benefits of  sEMG in implementation, Integrating sEMG findings with other diagnostic modalities can be challenging due to differences in terminology and interpretation. EMG is often 
used in conjunction with other diagnostic tests, such as nerve conduction studies or imaging, to provide a comprehensive assessment of neuromuscular function.

\subsection{Prior Solutions}
Over the years, attempts have been made to address the challenges associated with electromyography (sEMG) signal interpretation in clinical practice. These efforts have included the use of filtration 
techniques such as analog filters, but these have been found to be inefficient in addressing the complexities of sEMG recordings. Additionally, sEMG pattern variability has been addressed through 
classification methods that involve signal segmentation using convolutional neural networks (ConvNet). While this approach has shown promise, it has not fully resolved the issue of sEMG pattern variability. 
In the biomedical application of sEMG for prosthetic leg control, inadequate data filtration from analog filters has resulted in unclear movement of the prosthetic leg. However, one of the biggest 
challenges in the clinical application of sEMG is its use in the early diagnosis of neurological disorders. Despite its potential, sEMG can be limited in its ability to detect subtle changes in muscle 
activity that may indicate the presence of a neurological disorder in its early stages. Therefore, further research and innovation are needed to develop standardized norms, quality assurance measures, and 
patient-centered approaches for sEMG interpretation in the context of early diagnosis of neurological disorders. The objective of this study is to build on prior solutions and identify new approaches that 
can address the challenges associated with sEMG in clinical practice, particularly in the early diagnosis of neurological disorders.\\
These are last year prior soltuions for this project:
\begin{itemize}
    \item Previous attempts at solving sEMG signal interpretation challenges have utilized filtration techniques (analog filters) but these were found to be inefficient.
    \vspace{0.2cm}
    \item sEMG pattern variability has been addressed through classification methods involving signal segmentation using convolutional neural network (ConvNet), but this approach did not fully resolve 
    the issue.
    \vspace{0.2cm}
    \item In the biomedical application of sEMG for prosthetic leg control, inadequate data filtration from analog filters has resulted in unclear movement of the prosthetic leg.
  \end{itemize}
  Further research and innovation are needed to develop standardized norms, quality assurance measures, and patient-centered approaches for sEMG interpretation.

\subsection{Engineering Goals}
Electromyography (sEMG) is a technique widely used in clinical practice for the assessment of neuromuscular disorders. Despite its potential clinical benefits, sEMG recordings can be challenging to 
interpret due to the variability among individuals and muscles. Moreover, technical limitations in sEMG recordings can affect accuracy and reproducibility, which can compromise its reliability as a 
diagnostic tool. To address these challenges, this study aims to identify, compare, and implement solutions that can enhance the accuracy and reliability of sEMG recordings in clinical practice.
\begin{itemize}
    \item Develop standardized protocols for accurate sEMG signal interpretation. Establish normative values for sEMG patterns that account for variability among individuals and muscles.
    \vspace{0.2cm}
    \item Improve technical aspects of sEMG recordings to enhance accuracy and reproducibility. Explore strategies for integrating sEMG findings with other diagnostic modalities to improve 
    diagnostic consistency and patient care.    
 \end{itemize}
 Aiming to improve the accuracy and reliability of electromyography (sEMG) recordings in clinical practice for the assessment of neuromuscular disorders. By developing standardized protocols for sEMG 
 signal interpretation and improving technical aspects of sEMG recordings using filters and AI, this study seeks to enhance diagnostic consistency, controlling of prosthetic and patient care. It is 
 hoped that the results of this study will contribute to the ongoing efforts to improve the clinical application of sEMG and benefit patients with neuromuscular disorders.

\subsection{Challenges in the last solution}
Last soltuion, there significant challenges faced it, specifically in filtering signals using analog filter for biomedical application such controlling lower limb prosthetic leg or architecture of AI to train these signals. 

\begin{itemize}
    \item Although numerous attempts have been made to improve the filtration process in interpreting sEMG signals, especially when using analog filters, these efforts have not 
    been entirely successful. Residual noise has remained unfiltered, resulting in inaccuracies in the output signal. As a result, it is crucial to continue research to 
    optimize the filtration process and the develpoment of filters, particularly for filtering complex sEMG signals. Advanced signal processing techniques, such as digital filtering and wavelet 
    analysis, may hold promise for effectively filtering out unwanted noise while preserving the integrity of the sEMG signal. 
    
    \begin{figure}[h]
        \centering
        \includegraphics[width=0.4\textwidth]{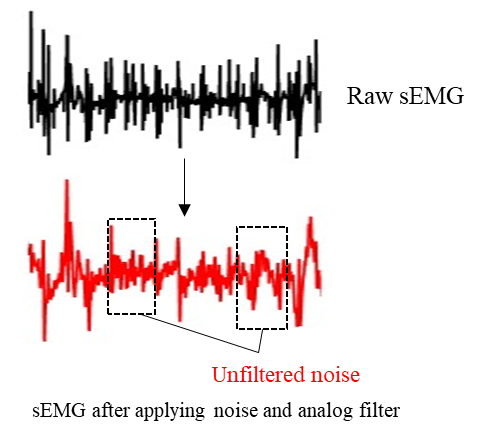}
        \caption{The first signal is a sEMG signal devoid of any extraneous noise. The second signal denotes a signal that has been subject to power line noise, followed 
        by analog filtering. While the signal has undergone filtration, there persists residual noise that remains unfiltered.}
        \label{fig:figure1}
    \end{figure}

      \vspace{0.2cm}
    
      \item Convolutional networkds (ConvNets) face significant challenges in achieving real-time processing and high classification accuracy rates when applied to bio-signals due to signal and 
      pattern similarity, as well as parameter overload. To overcome these challenges, a customized ConvNet architecture should be developed to achieve real-time processing 
      and high accuracy classification.
      \begin{figure}[h]
        \centering
        \includegraphics[width=0.4\textwidth]{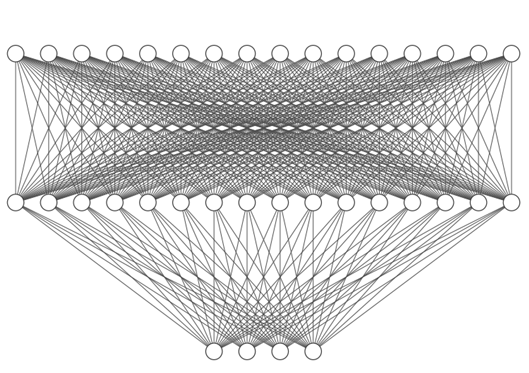}
        \caption{Fully Connected (FC) layer in neural networks can improve the accuracy of each node and weight in the model. However, this approach can lead to a 
        deeper architecture and longer processing times due to the increased complexity of the network.}
        \label{fig:figure1}
      \end{figure}

      \vspace{0.2cm}

      \item In order to effectively integrate prosthetic leg with sEMG technology, it is crucial to address Challenge 1 and Challenge 2. The presence of noise in 
      the sEMG signal can significantly impact the accuracy of the classification process, thereby affecting the overall movement of the prosthetic leg.
      \begin{figure}[h]
        \centering
        \includegraphics[width=0.4\textwidth]{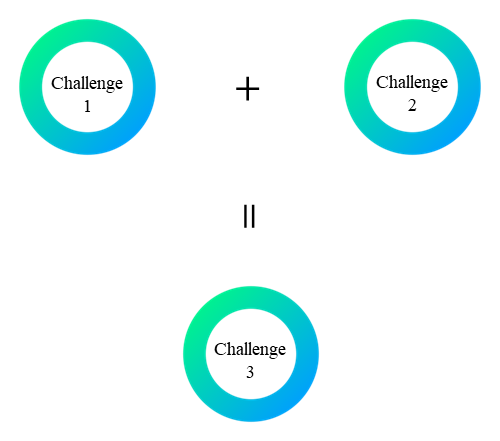}
        \label{fig:figure1}
      \end{figure}

\end{itemize}

\section{Methodology}

\subsection{Analog vs Digital Filters}

Analog filters are constructed by employing analog circuitry, which involves the use of hardware components. On the other hand, digital filters are designed utilizing digital signal processing techniques, which rely on software implementation. The process of designing analog filters typically entails the careful selection and connection of passive or active components within the circuitry. In contrast, digital filters are based on mathematical algorithms that are implemented in software.
One notable advantage of digital filters is their ability to offer greater flexibility and precision in adjusting filter characteristics compared to analog filters. This flexibility stems from the inherent versatility of software-based algorithms, allowing for fine-tuning of various filter parameters. In contrast, analog filters may exhibit limitations in terms of flexibility and precision due to the constraints imposed by physical circuit components.
\begin{figure}[h]
    \centering
    \includegraphics[width=0.5\textwidth]{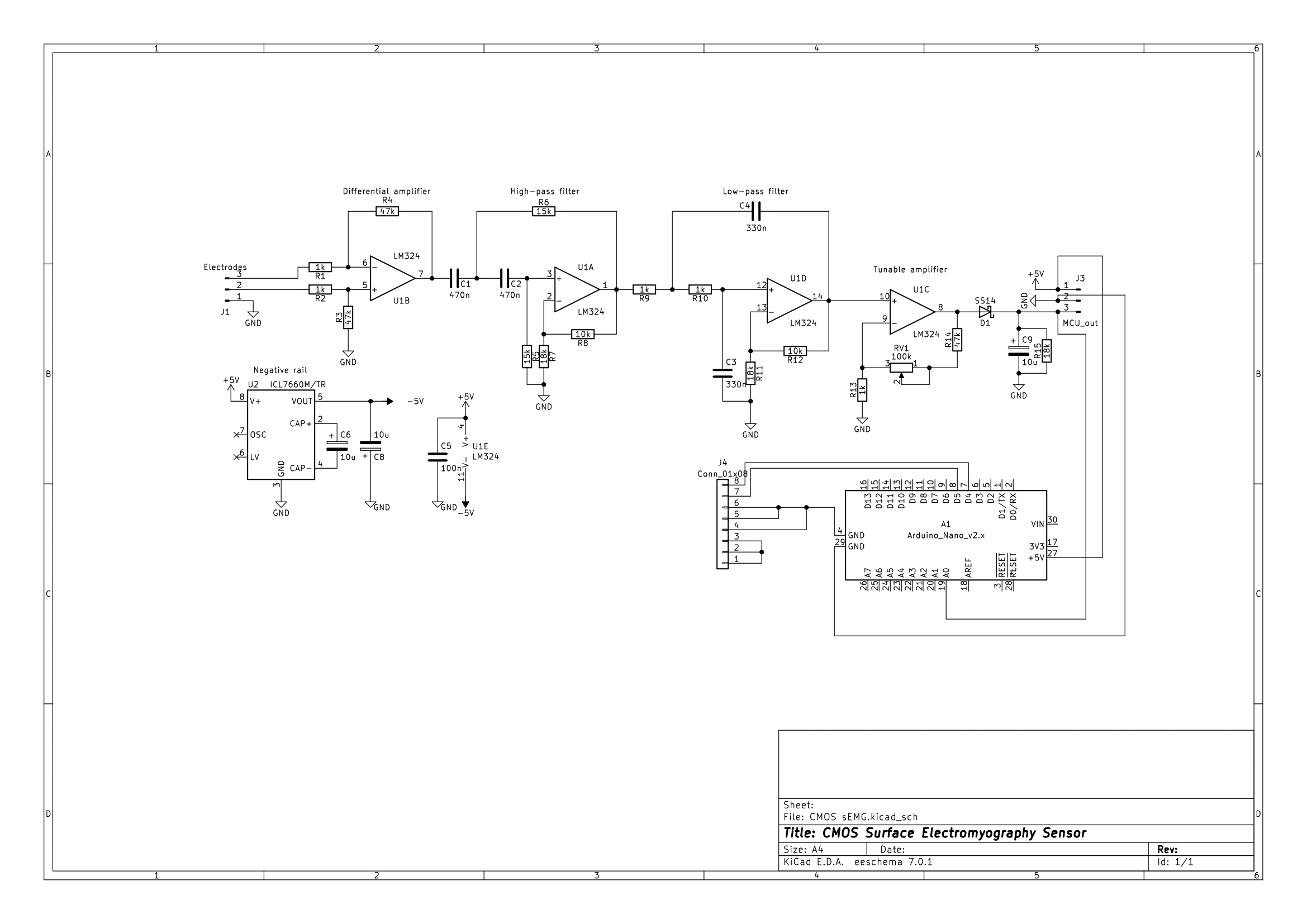}
    \caption{A schematic diagram for sEMG with the implementation of analog filters in a surface electromyography (sEMG) system. Analog filters play a crucial role in filtering and processing the sEMG signals.}
  \end{figure}

\subsection{Noise sensitivity in Analog Vs. Digital filters}

\textbf{Analog:}

\begin{itemize}
  \item Susceptible to environmental factors and component tolerances, which can introduce noise.
  \item Prone to external noises (e.g., EMI, RFI) that can degrade signal-to-noise ratio (SNR).
\end{itemize}

\textbf{Digital:}

\begin{itemize}
  \item Less susceptible to environmental factors and component tolerances.
  \item Precise control over filter characteristics, allowing for high accuracy and predictable performance.
  \item Can mitigate external noises.
\end{itemize}

\begin{figure}[h]
    \centering
    \includegraphics[width=0.45\textwidth]{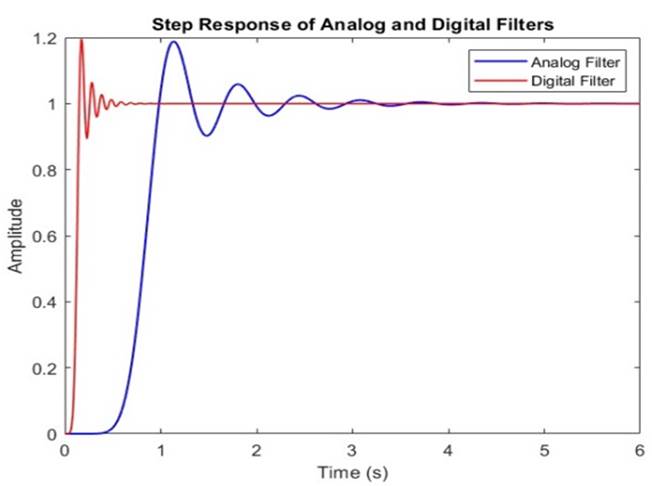}
    \caption{The step response refers to the output response and provide information about the system when a sudden unit step input is applied.}
  \end{figure}
\begin{figure}[h]
    \centering
    \includegraphics[width=0.5\textwidth]{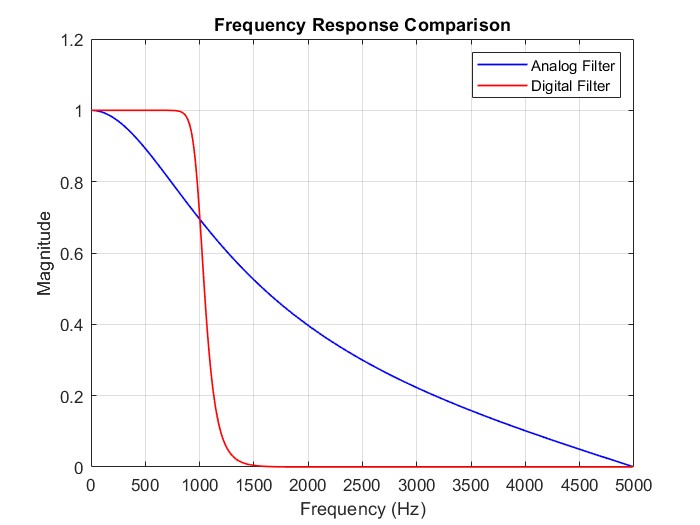}
    \caption{Frequency response describes how a system or device responds to different frequencies of an input signal after applying cutoff frequency.}
  \end{figure}
Based on the previous discussion, it can be inferred that digital filters generally offer advantages over analog filters in terms of flexibility, adaptability, and stability. In the next section, we will delve into the specific comparison of the output characteristics of digital and analog filters in the context of noise filtering.

\subsection{Testing of Analog VS Digital Filters noise filtering}
The process of testing was as following: 
\begin{enumerate}
    \item A comparative study was conducted to evaluate the performance of digital and analog filters in filtering out noise from a given input signal.
    \item A test signal contaminated with noise was generated, representing a real-world scenario where a signal of interest is corrupted by unwanted noise.
    \item Both digital and analog filters were designed to have similar filter characteristics, such as cutoff frequency, filter order, and filter type, to ensure a fair comparison.
    \item The input signal contaminated with noise was passed through both filters separately, and the filtered output signals were recorded for further analysis.
\end{enumerate}
\begin{figure}[h]
    \centering
    \includegraphics[width=0.45\textwidth]{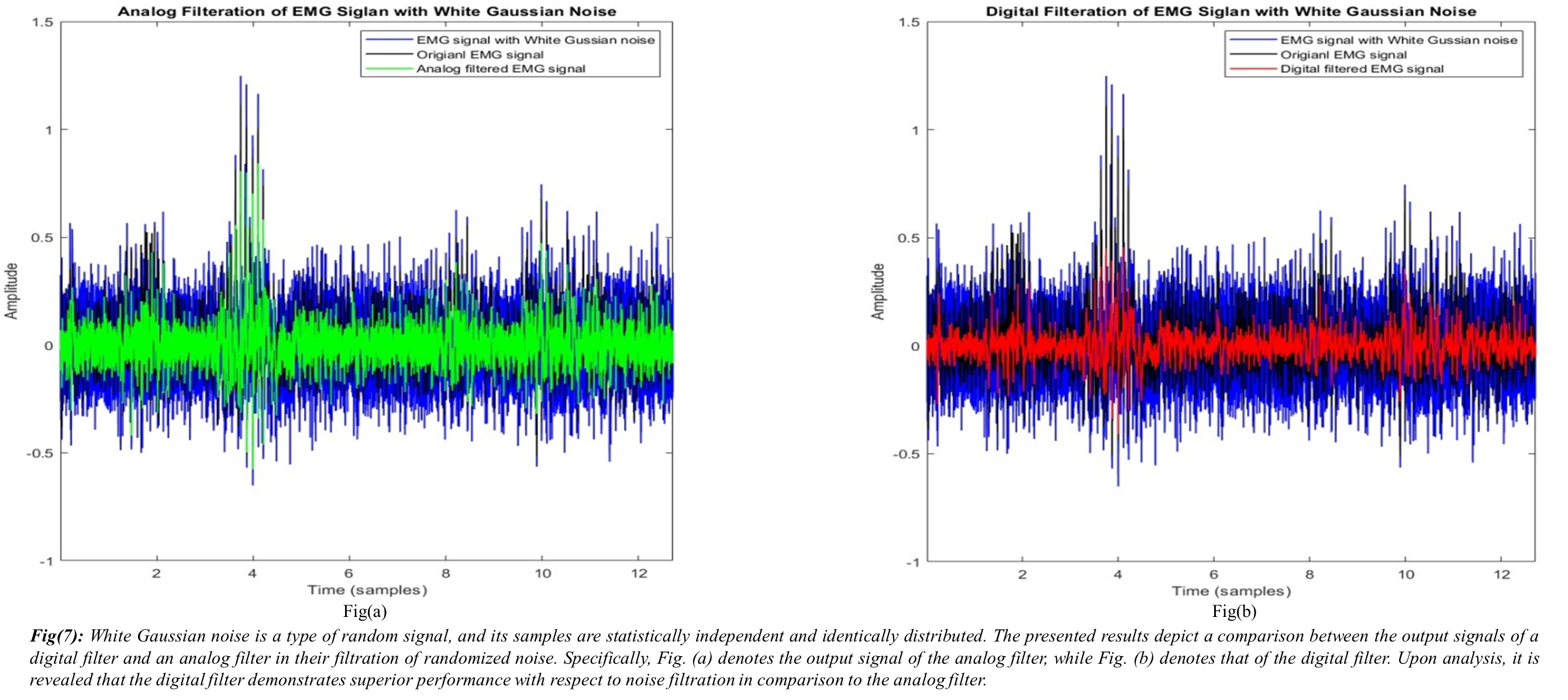}
    \caption{White Gaussian noise is a type of random signal and its samples are statistically independent and identically distributed (i.i.d.), which means that each sample is unrelated to the previous or subsequent samples. The presented results depict a comparison between the output signals of a digital filter and an analog filter in their filtration of randomized noise. Specifically, Fig. (a) denotes the output signal of the digital filter, while Fig. (b) denotes that of the analog filter. Upon analysis, it is revealed that the digital filter demonstrates superior performance with respect to noise filtration in comparison to the analog filter.}
\end{figure}

\begin{figure}[h]
    \centering
    \includegraphics[width=0.4\textwidth]{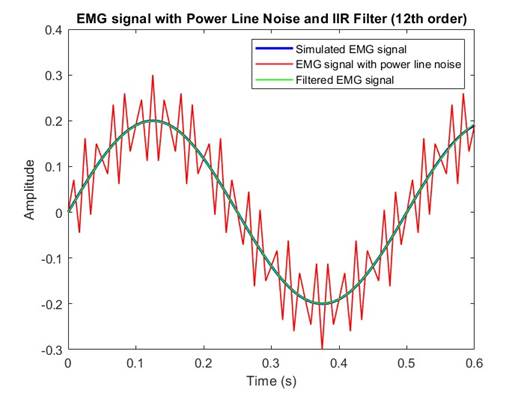}
    \caption{Power line noise, refers to unwanted electrical disturbances that can be present electrical systems. Power line noise can manifest as electrical signals with frequencies that are superimposed on the normal 50 or 60 Hz waveform.}
\end{figure}
    
\begin{figure}[h]
    \centering
    \includegraphics[width=0.4\textwidth]{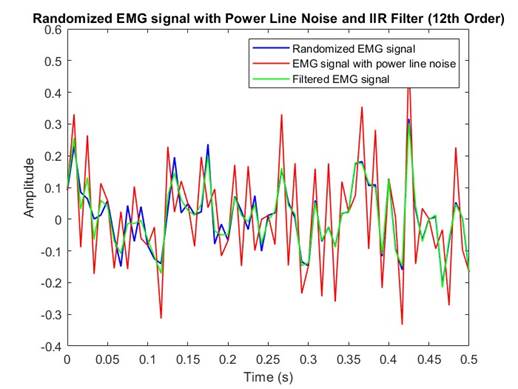}
    \caption{Randomized EMG signals were subjected to filtration through an IIR digital filter. The aim of the experiment was testing the stability response and evaluate the filtration accuracy of the applied filter.}
\end{figure}

\subsection{Optimizing ConvNet for real-time signal processing}
Comparing the accuracy of ResNet18 and VGG16 for classification; using VGG16 and ResNet18 for signal processing and classification on different channels (15 subjects' public EEG and 18 observed EMG data), based on that ConvNets are compared on different signals and channels as shown in fig(9).
Signals are preprocessed and A frequency filter ranging from 0 to 75 Hz was used to apply a band pass to the signal at 200 Hz for more significant analysis.

\begin{figure}[h]
    \centering
    \includegraphics[width=0.4\textwidth]{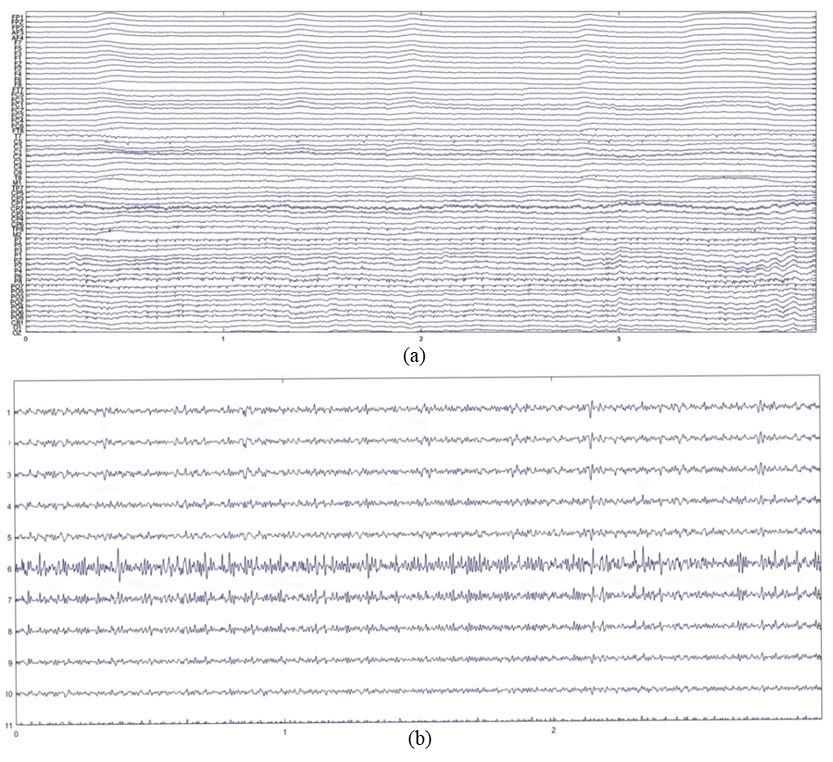}
    \caption{A representation of collected EEG signal before preprocessing from all 62 electrodes and 10 ones to compare between its complexity in training using VGG16 and ResNet as (a) illustrates the collected 62-channels and (b) demonstrates 10 channels.}
\end{figure}
we conducted a performance comparison between VGG16 and ResNet models for EEG classification. Specifically, we examined the impact of VGG16's fully connected layers and deeper architecture on the classification performance in a 62-channels EEG dataset, in comparison to ResNet. Surprisingly, despite VGG16 demonstrating higher performance in a 10-channel EEG dataset, our findings revealed that it exhibited lower performance in the more complex 62-channels EEG dataset. These results suggest that the fully connected layers and deeper architecture of VGG16 may not be as effective in capturing the intricate patterns present in the 62-channels EEG data, in contrast to the superior performance of ResNet.

\begin{figure}[h]
    \centering
    \includegraphics[width=0.4\textwidth]{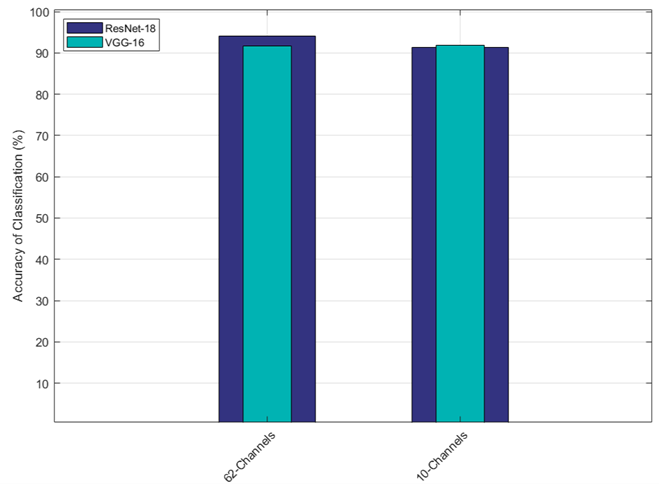}
    \caption{Performance Comparison of VGG16 and ResNet in EEG Classification: VGG16's Fully Connected Layers and Deeper Architecture May Contribute to Lower Performance in 62-Channels EEG, Despite Higher Performance in 10-Channel EEG, Compared to ResNet.}
\end{figure}

\begin{figure*}[b]
    \centering
    \includegraphics[width=\textwidth]{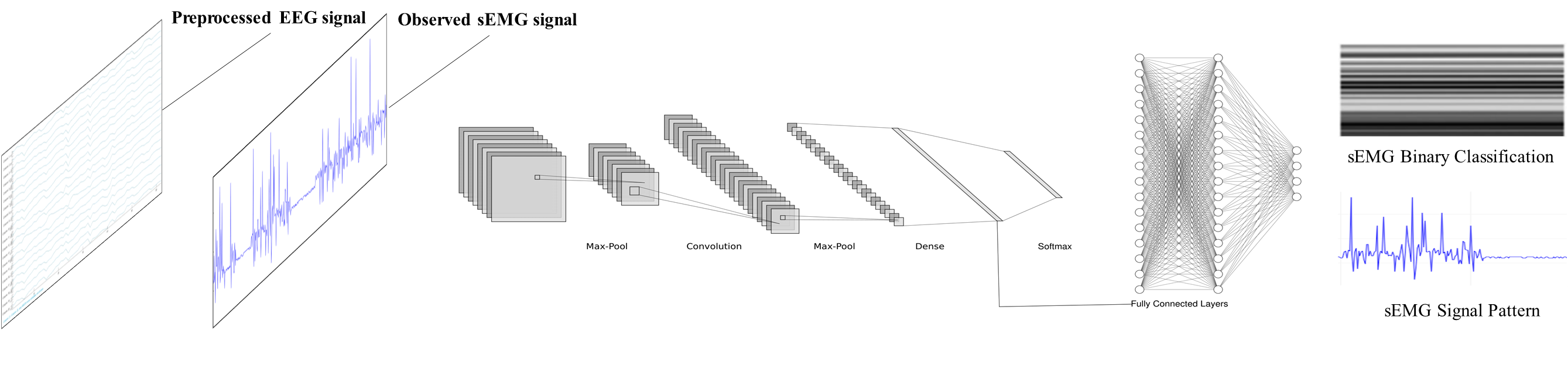}
    \caption{In this study, a customized VGG architecture was developed to mitigate the reduced accuracy of VGG16 in training big data of EEG signals on different channels compared to ResNet. The customized VGG architecture was designed to be less complex and reduce parameters, aiming to optimize its performance. Comparative analysis revealed that the customized VGG architecture exhibited improved accuracy compared to VGG16, providing valuable insights for researchers and practitioners in the field of deep learning for Bio-signals classification. In depth, The Customized architecture was developed by utilizing small (1, 3) convolutional kernels and removing fully connected (FC) layers to enhance the model, reduce parameters and avoid overtraining and over fitting.}
  \end{figure*}

  \begin{figure*}[b]
    \centering
    \includegraphics[width=\textwidth]{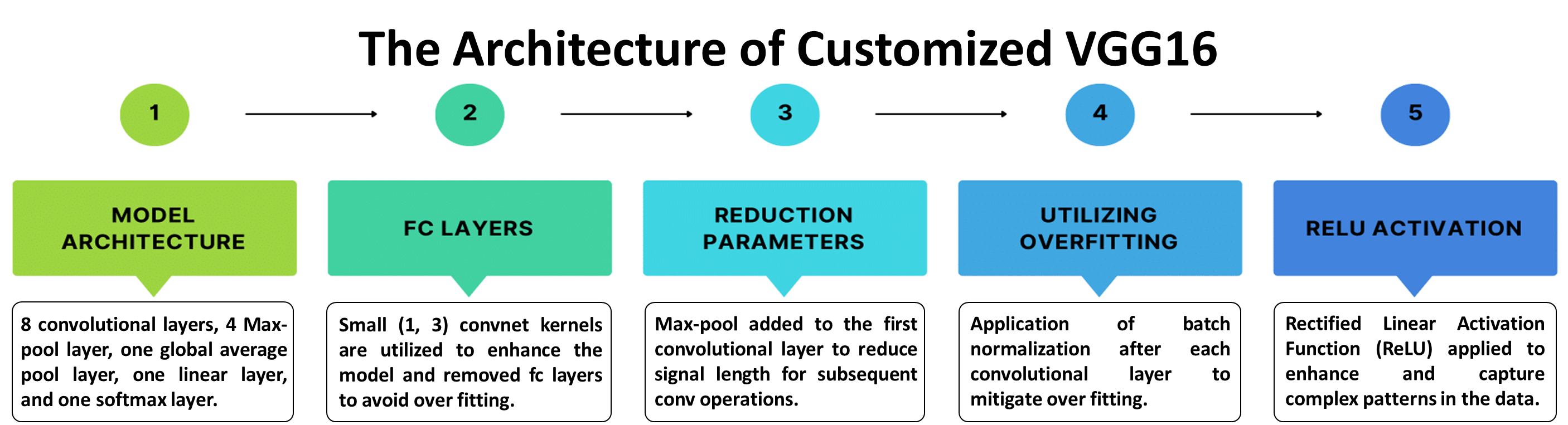}
  \end{figure*}
we aimed to address the reduced accuracy of VGG16 when training large-scale EEG datasets with varying channel configurations, in comparison to ResNet. To mitigate this issue, we developed a customized VGG architecture specifically tailored for EEG signal classification. The customized VGG architecture was designed to be less complex, optimizing its performance by reducing the number of parameters.
Our comparative analysis revealed that the customized VGG architecture outperformed VGG16 in terms of accuracy, showcasing its effectiveness in handling the challenges posed by diverse EEG signal datasets. This finding holds great significance for researchers and practitioners in the field of deep learning for bio-signal classification.
To enhance the customized architecture, we employed small convolutional kernels (1, 3) dimensions and eliminated fully connected (FC) layers. These modifications aimed to improve the model's capacity, reduce the number of parameters, and prevent overtraining and overfitting issues commonly encountered in deep learning models.
The development of this customized VGG architecture demonstrates an innovative and scientific approach to tackling and imrpove the performance and the specific challenges associated with EEG signal classification. The results highlight the potential for optimized deep learning models tailored for bio-signal analysis, contributing valuable insights to the scientific community and opening up new avenues for future research.

\subsection{Optimizers and Hyperparameters Art}
Choosing between ADAMS and SGD as VGG optimizer:
\begin{enumerate}
    \item Investigated the performance of two optimization algorithms, SGD and Adam, on the VGG16 model for deep learning tasks.
    \item Conducted experiments with both SGD and Adam optimizers and compared their performance in terms of model accuracy. 
    \item Our findings revealed that Adam optimizer outperformed SGD in terms of accuracy, achieving higher accuracy rates on the validation and test sets. ADAMS avoided overtraining.
    \end{enumerate}
\begin{figure}[h]
        \centering
        \includegraphics[width=0.5\textwidth]{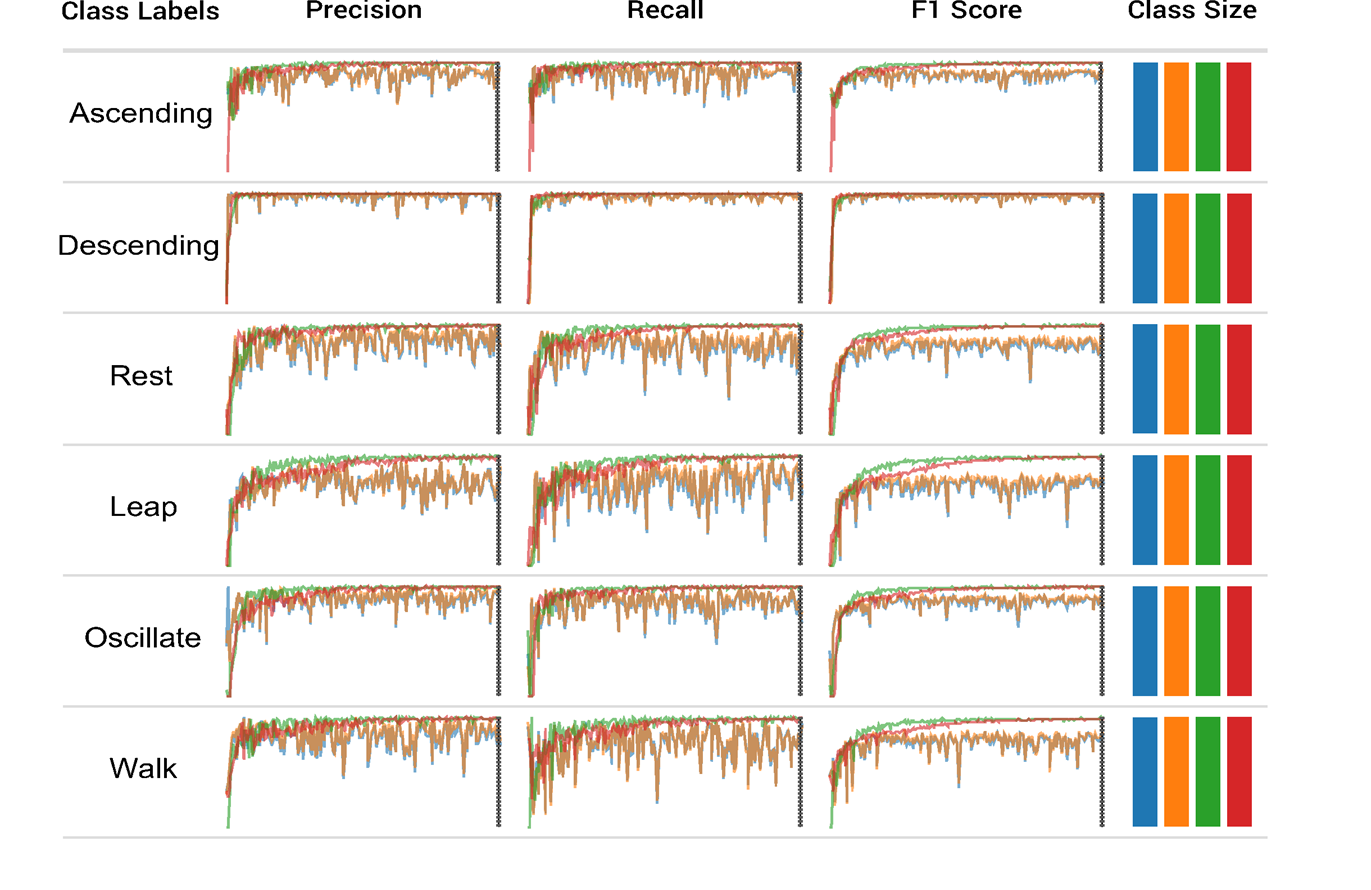}
\end{figure}

\begin{figure}[h]
    \centering
    \includegraphics[width=0.45\textwidth]{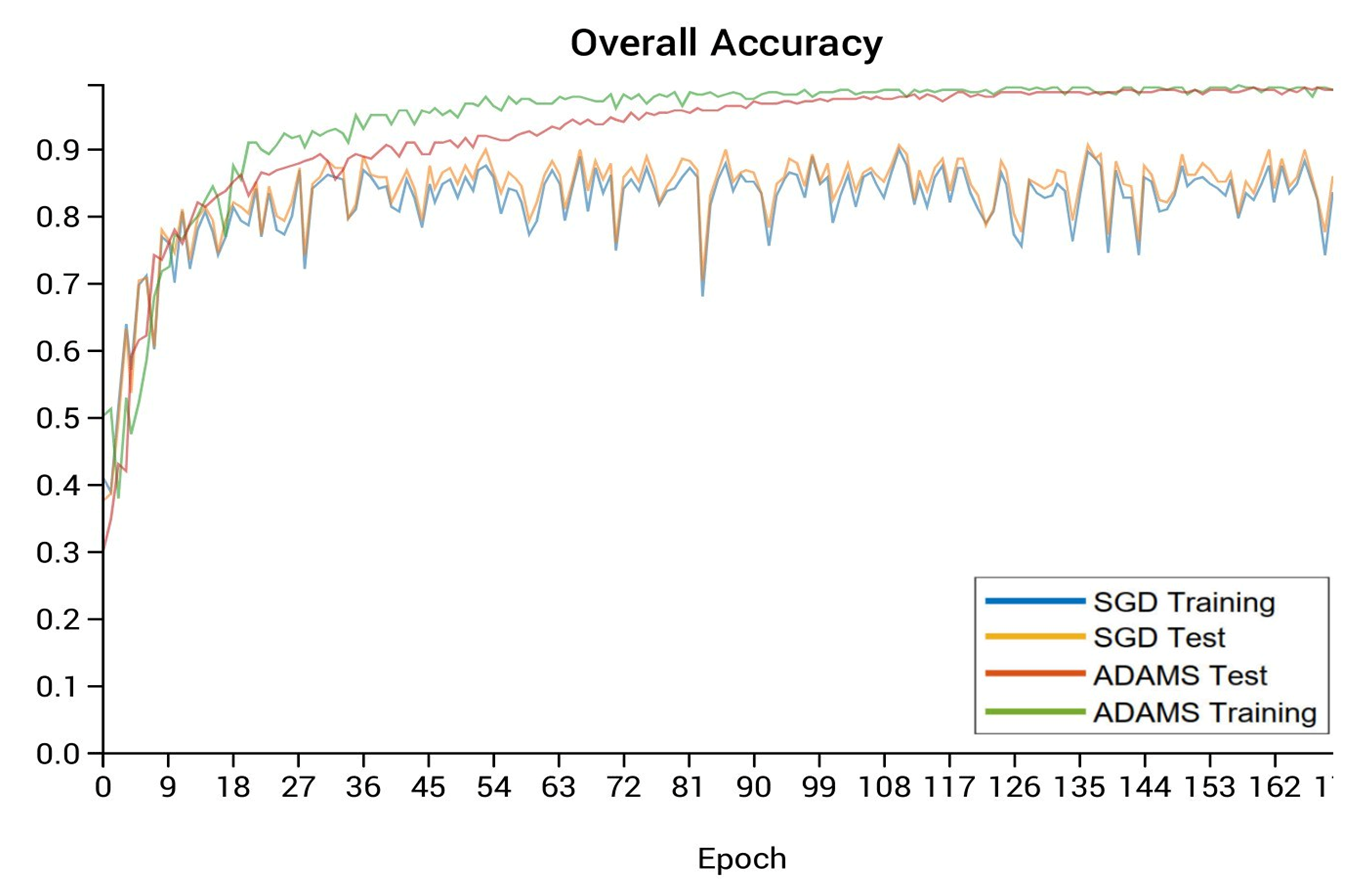}
    \caption{Optimizing Deep Learning: A Comparative Study of SGD and Adam Optimizers for VGG16 Model. Adams Chosen for Overtraining Avoidance and Higher Accuracy over SGD. Optimizing VGG Model Training and Validation for Time Series and Image Data using PyTorch 1.7.0 with CUDA 11: Investigating the Impact of Hyper parameters, including Learning Rate (1e-3), Epochs (200), and Optimizers (Adams and SGD as ADAMS did an efficient impact and avoided overtraining than SGD).}
\end{figure}
In the pursuit of optimizing deep learning models, a comprehensive study was conducted to compare the effectiveness of two popular optimizers, namely SGD and Adam, in the context of the VGG16 model. Adam emerged as the preferred choice due to its ability to prevent overtraining and deliver superior accuracy, surpassing the performance of SGD. This investigation sheds light on the selection of optimizers.

\section{Data Analysis and Results}
Two approaches in signal processing:
\begin{itemize}
    \item Performing VGG-16 1-Dimensional convolutions, which are applied along a time axis to capture patterns in the time domain.
    \item Performing VGG-16 2-Dimensional convolutions, which are applied in both X and Y dimensions to capture patterns in images.
\end{itemize}
\begin{figure}[h]
    \centering
    \includegraphics[width=0.5\textwidth]{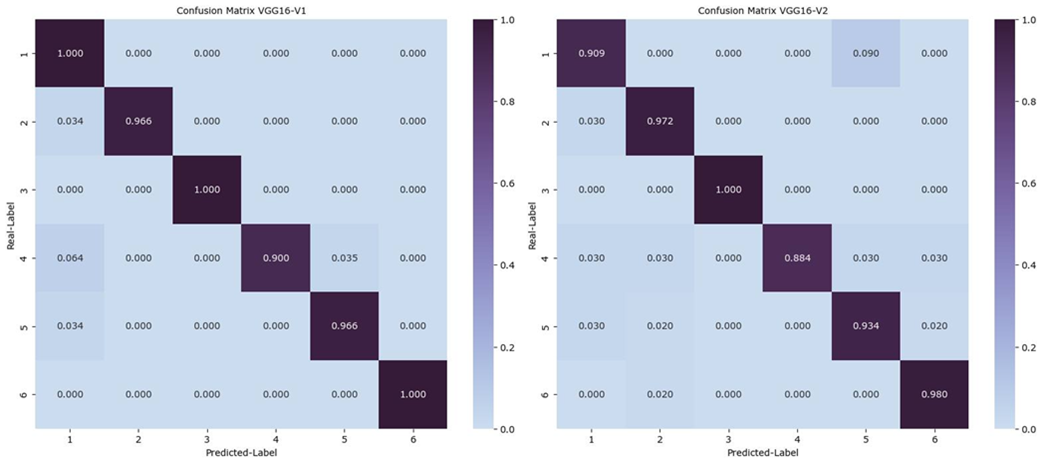}
    \caption{Finding the recognition rate for each class as 1:6 represent Ascending, Descending, Resting, Leaping, Oscillating and Walking. To compare between VGG16-V1 and VGG16-V2 recognition rate. }
\end{figure}

\begin{figure}[h]
    \centering
    \includegraphics[width=0.5\textwidth]{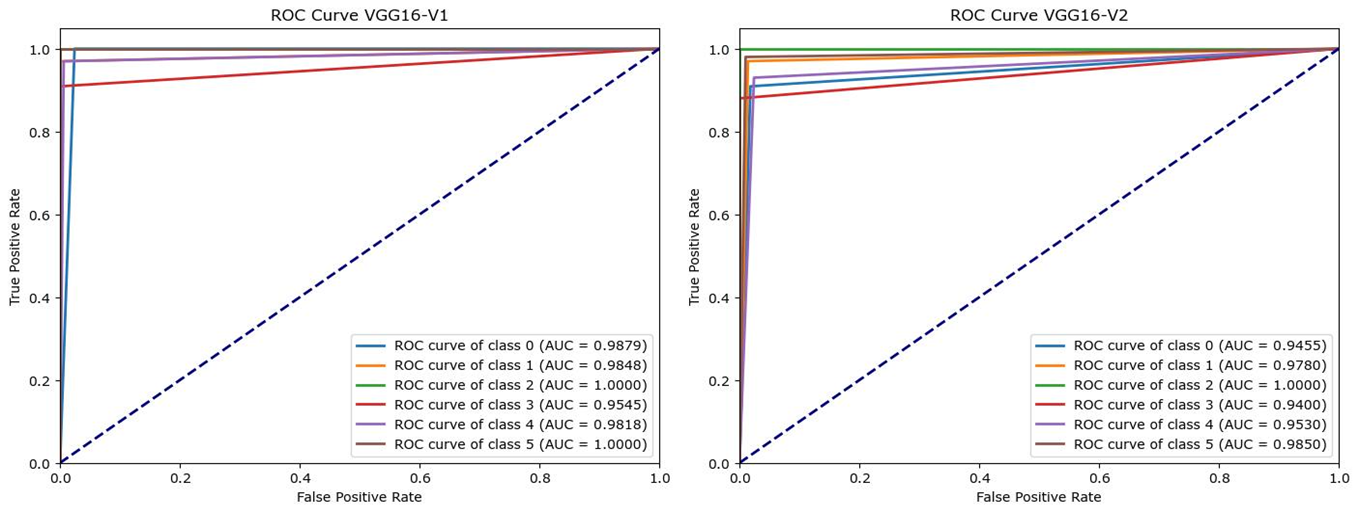}
    \caption{Receiver Operating Characteristic curve (ROC) is showing the performance of a classification model at all classification thresholds as ]0.5 : 1.0] indicates for significant classification.}
\end{figure}

\begin{figure}[h]
    \centering
    \includegraphics[width=0.5\textwidth]{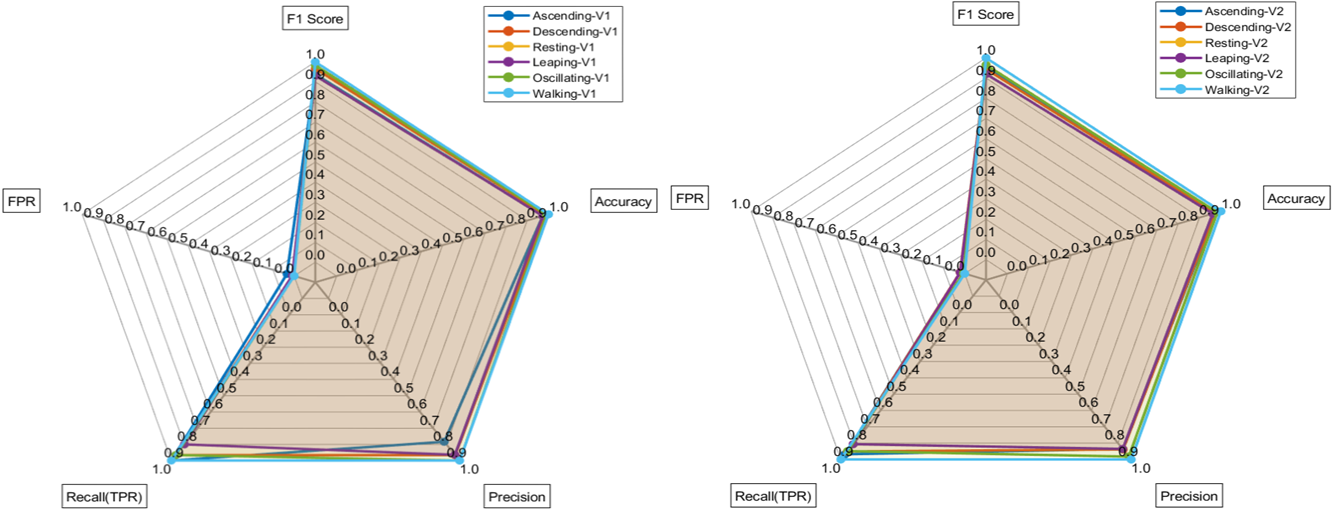}
    \caption{Comparing between one and two dimensional kernels VGG16 in multi-classification (six-class) in 5 aspects: F1 Score, Accuracy, Precision, Recall (True Positive  Rate), and False Positive Rate (FPR).}
\end{figure}

The pattern-based approach of VGG16-V1 provided and emphasis on capturing intricate patterns, achieved a higher recognition rate and accuracy compared to VGG16-V2, which employed a binary approach. VGG16-V1's architecture was specifically designed to learn and recognize complex patterns by utilizing a deeper network structure with multiple convolutional layers compared to the binary approach of VGG16-V2.

A randomized analysis of electromyography (EMG) signals and examined the stability of an Infinite Impulse Response (IIR) filter in the presence of power line noise. The objective was to investigate the deficiencies that can occur in analog filters when filtering EMG signals corrupted by power line interference.

\begin{figure}[h]
    \centering
    \includegraphics[width=0.38\textwidth]{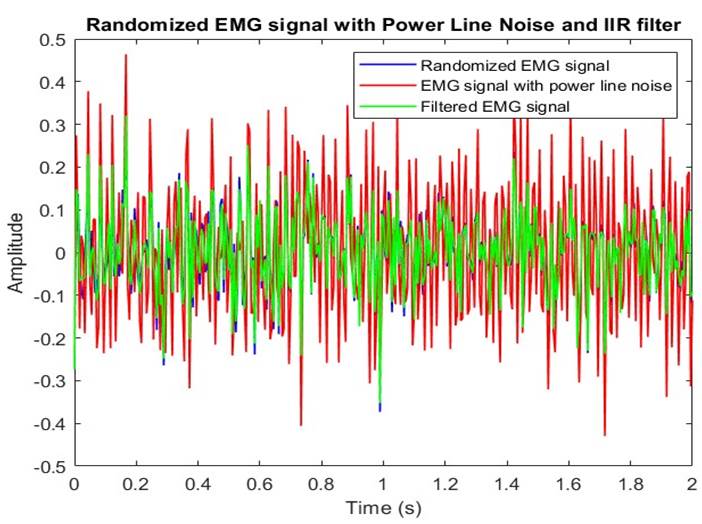}
    \caption{Randomized EMG signal and testing the stability of the IIR filter with power line noise to see the lack which happened in the analog filter.}
\end{figure}

\begin{figure}[h]
    \centering
    \includegraphics[width=0.38\textwidth]{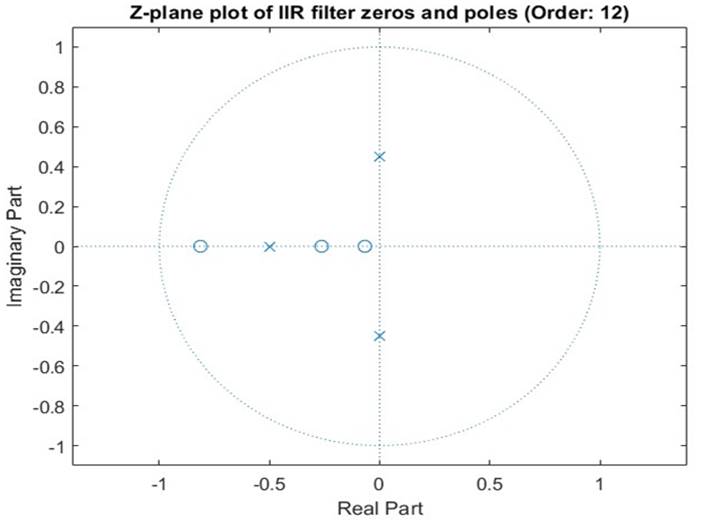}
    \caption{A Comprehensive Visualization for IIR filter in z-plane, which shows its stability.}
\end{figure}

\begin{figure}[h]
    \centering
    \includegraphics[width=0.38\textwidth]{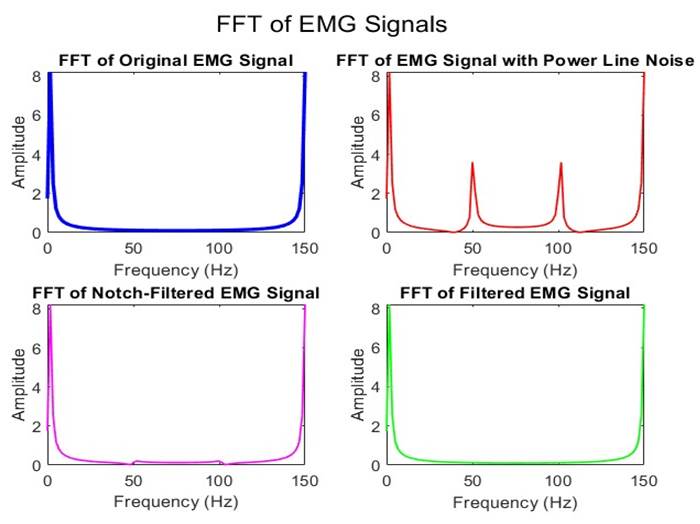}
    \caption{Demonstration on how noise being removed from the signal on FFT (using notch-filter).}
\end{figure}

\begin{table}[htbp]
    \centering
    \caption{Comparative analysis of analog and digital filters in terms of their Mean Squared Error (MSE) and Root Mean Squared Error (RMSE) performance metrics.}
    \label{tab:filter_comparison}
    \begin{tabular}{|c|c|c|}
    \hline
    \textbf{Filter Type} & \textbf{MSE} & \textbf{RMSE} \\ \hline
    Analog Filter        & 0.0047       & 0.0686        \\ \hline
    Digital Filter       & 0.0011       & 0.0327        \\ \hline
    \end{tabular}
    \end{table}

\begin{table}[htbp]
    \centering
    \caption{Comparing VGG16V1 and VGG16-V2: A Comprehensive Analysis of Performance, Efficiency, and Computational Overhead for classification in muscle activities.    }
    \label{tab:comparison}
    \begin{tabular}{|c|c|c|}
    \hline
    \textbf{} & \textbf{VGG16-V1} & \textbf{VGG16-V2} \\ \hline
    \textbf{Preprocessing Time (s)} & 0.003135 & 0.003345 \\ \hline
    \textbf{CPU Processing Time (s)} & 0.013343 & 0.95740 \\ \hline
    \textbf{GPU Processing Time (s)} & 0.002808 & 0.020899 \\ \hline
    \textbf{Model Parameters} & 784,803 & 67,146,211 \\ \hline
    \textbf{Accuracy (\%)} & 98.4383 & 97.1966 \\ \hline
    \textbf{Macro-F1} & 0.97 & 0.95 \\ \hline
    \end{tabular}
\end{table}
    
Based on the provided data in table (1, 2), VGG16-V1 is better than VGG16-V2 in several aspects. VGG16-V1 demonstrated higher pretreatment time, CPU inference time, GPU reasoning time, and had fewer model parameters (small segments in one length) compared to VGG16-V2. Additionally, VGG16-V1 achieved a higher accuracy and a slightly more accurate macro F1 score compared to VGG16-V2. These findings suggest that VGG16-V1 may be a more efficient and effective choice for the given task, potentially offering better performance with lower computational overhead. The application of digital filters, alongside a customized convolutional neural network (CNN) architecture, specifically the VGG16-V1 model, has enabled real-time processing of signals with enhanced filtering capabilities. This advancement has opened new avenues for improved readings and accurate classification bio-signal for biomedical applications, such a novel controlled prosthetic leg and early diagnosis of dystonia.
\\

\textbf{Optimizing Prosthetic Control: Leveraging Signal-based Statistics for Accurate Prediction of Movement Patterns}
\begin{itemize}
    \item Ascending: Achieved perfect accuracy of 100\% (1.000), indicating all instances of this class were correctly predicted as "Ascending".
    \item Descending: Achieved high accuracy of 96.6\% (0.966), with 3.4\% of instances misclassified as other classes.
    \item Resting: Achieved perfect accuracy of 100\% (1.000), demonstrating all instances of this class were correctly predicted as "Resting".
    \item Leaping: Achieved accuracy of 90\%, with 6.4\% of instances misclassified as others and 3.5\% of instances misclassified as "Walking".
    \item Walking: Achieved accuracy of 96.6\% (0.966), with 3.4\% of instances misclassified as other classes.
    \item Resting: Achieved perfect accuracy of 100\% (1.000), showing all instances of this class were correctly predicted as "Resting".
    
\end{itemize}

\section{Applications}
\subsection{Prosthetic Leg Model}
Making a prosthetic limb with a high bearing capacity,
flexibility, comfort, and shock absorption for long-term usage
requires considerable effort. When fabricating a prosthetic
limb, it should be lightweight for ease of control and have
a good load bearing capability. The prosthetic limb is constructed 
from lightweight but robust materials. The limb may
or may not have functioning knee and ankle joints, depending
on the site of the amputation. The socket is a very accurate
cast of your residual limb that fits snugly over it. It assists
in the prosthetic leg’s attachment to your body. Suspension
systems are used to secure the prosthesis, whether by sleeve
suction, vacuum suspension/suction, or distal locking by pin
or lanyard. As shown in fig (19), numbers of models that are
created for achieving engineering goals such as high bearing
capacity, light, and long-term use.
\begin{figure}[h]
    \centering
    \includegraphics[width=0.4\textwidth]{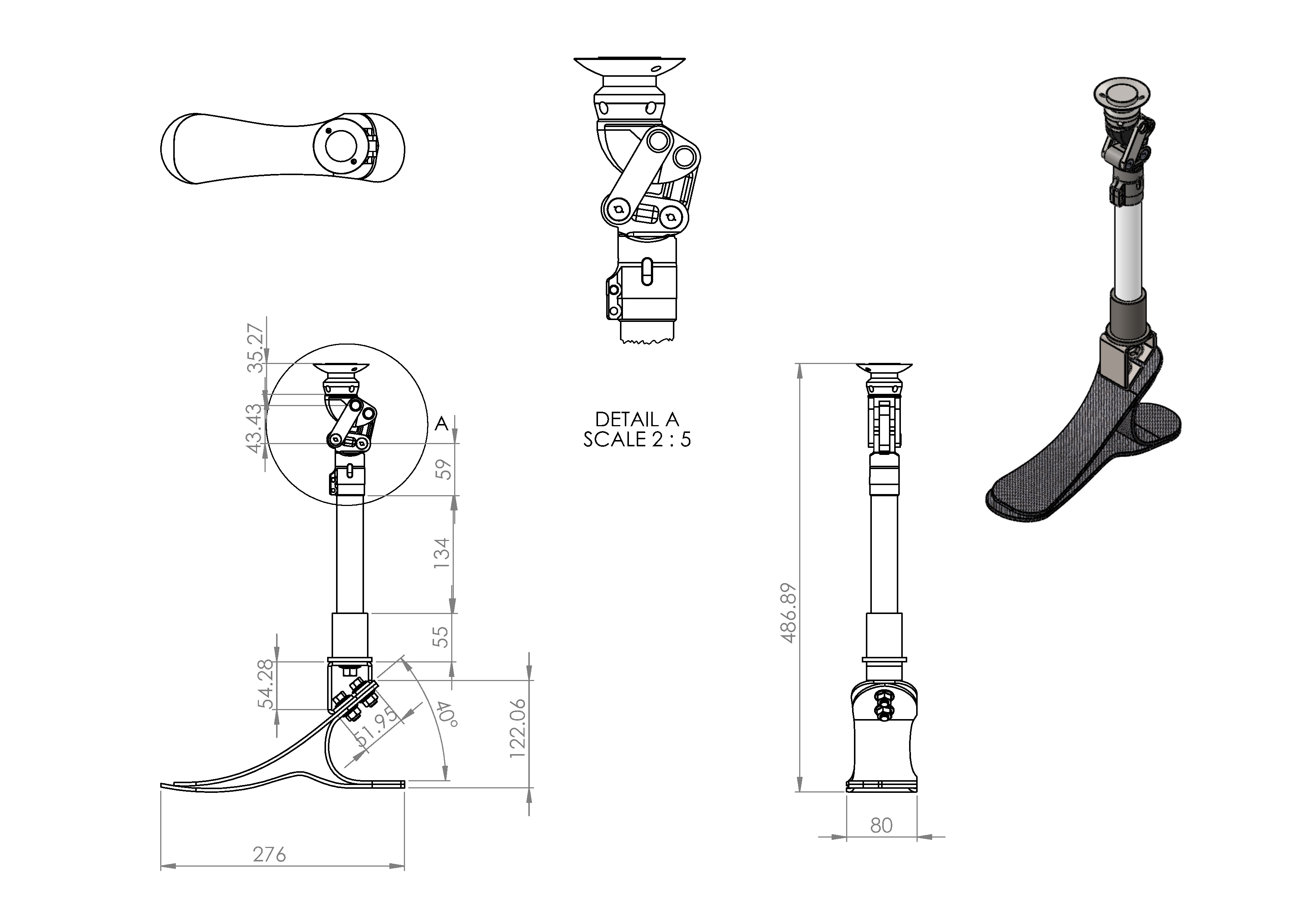}
    \caption{ 3D design of prosthetic limb.}
\end{figure}
\begin{figure}[h]
    \centering
    \includegraphics[width=0.4\textwidth]{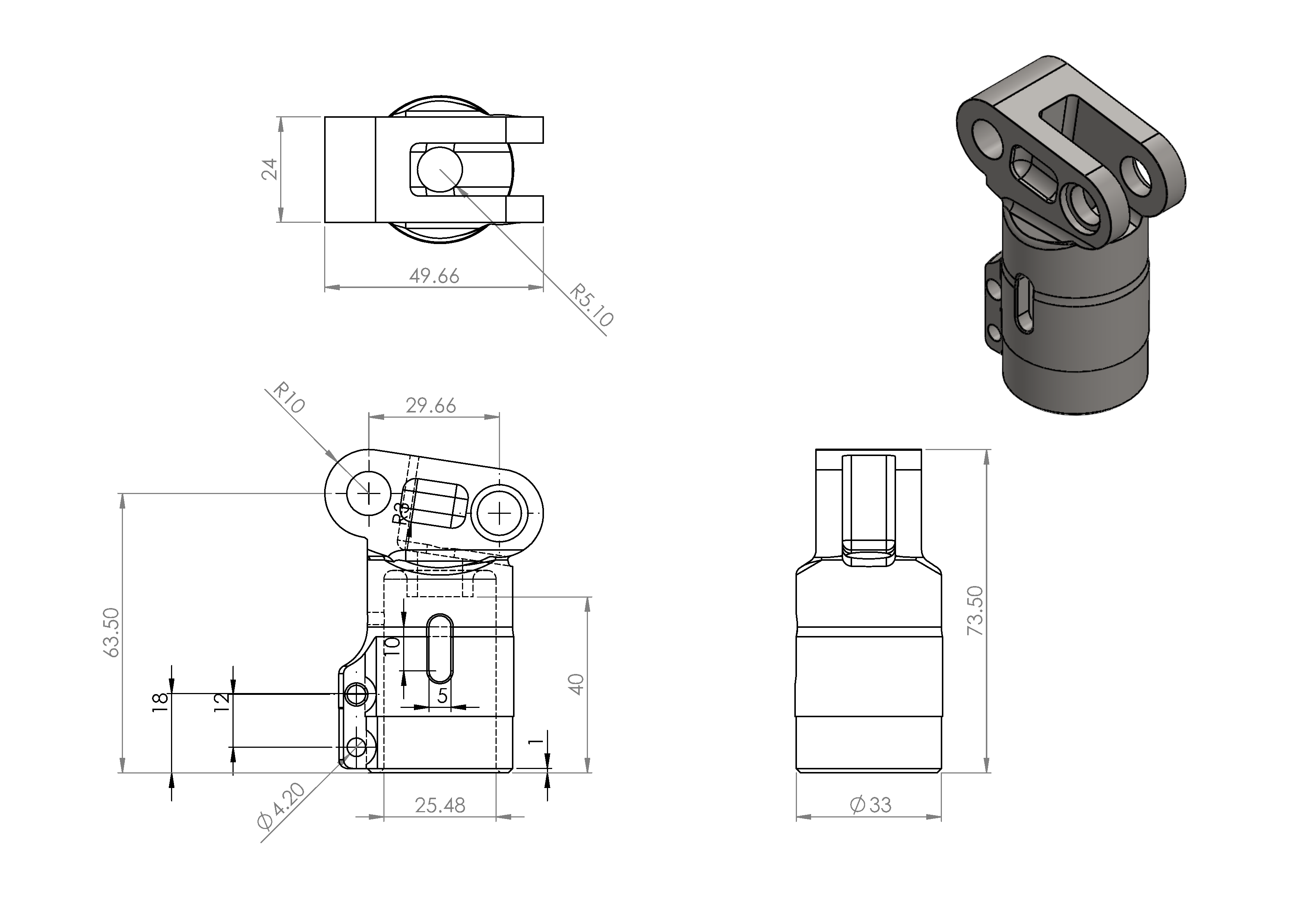}
    \caption{The design of prosthetic knee.}
\end{figure}
Following the selection of your prosthetic leg’s components,
you will need rehabilitation to strengthen your legs, arms,
and cardiovascular system as you learn to walk with your
new limb. You’ll work closely with rehabilitation experts,
physical therapists, and occupational therapists to develop a
rehabilitation plan that is customised to your unique mobility
requirements. Maintaining a healthy leg is a critical
component of this routine.
\begin{figure}[h]
    \centering
    \includegraphics[width=0.45\textwidth]{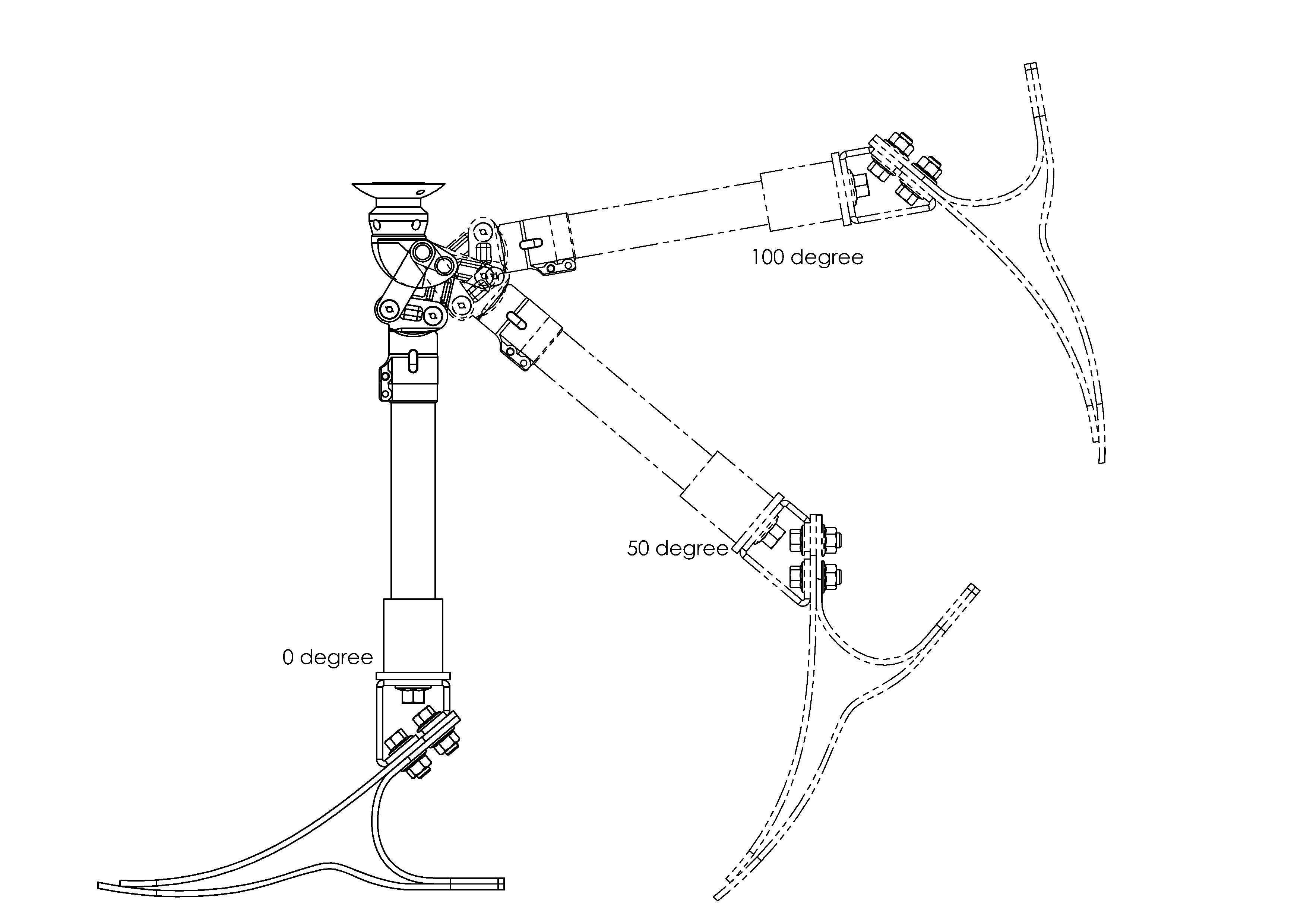}
    \caption{Real-life ration modeling for the prosthetic leg with 100 angel
    of movement.}
\end{figure}
In the field of prosthetics, surface electromyography (sEMG) has emerged as a promising technique for enhancing the control of prosthetic limbs. By capturing the electrical signals generated by the user's muscles through surface electrodes, sEMG enables the intuitive and precise control of prosthetic devices. This figure illustrates a prosthetic leg controlled by surface EMG, showcasing how it can be optimized to improve the user experience.

\begin{figure}[h]
    \centering
    \includegraphics[width=0.4\textwidth]{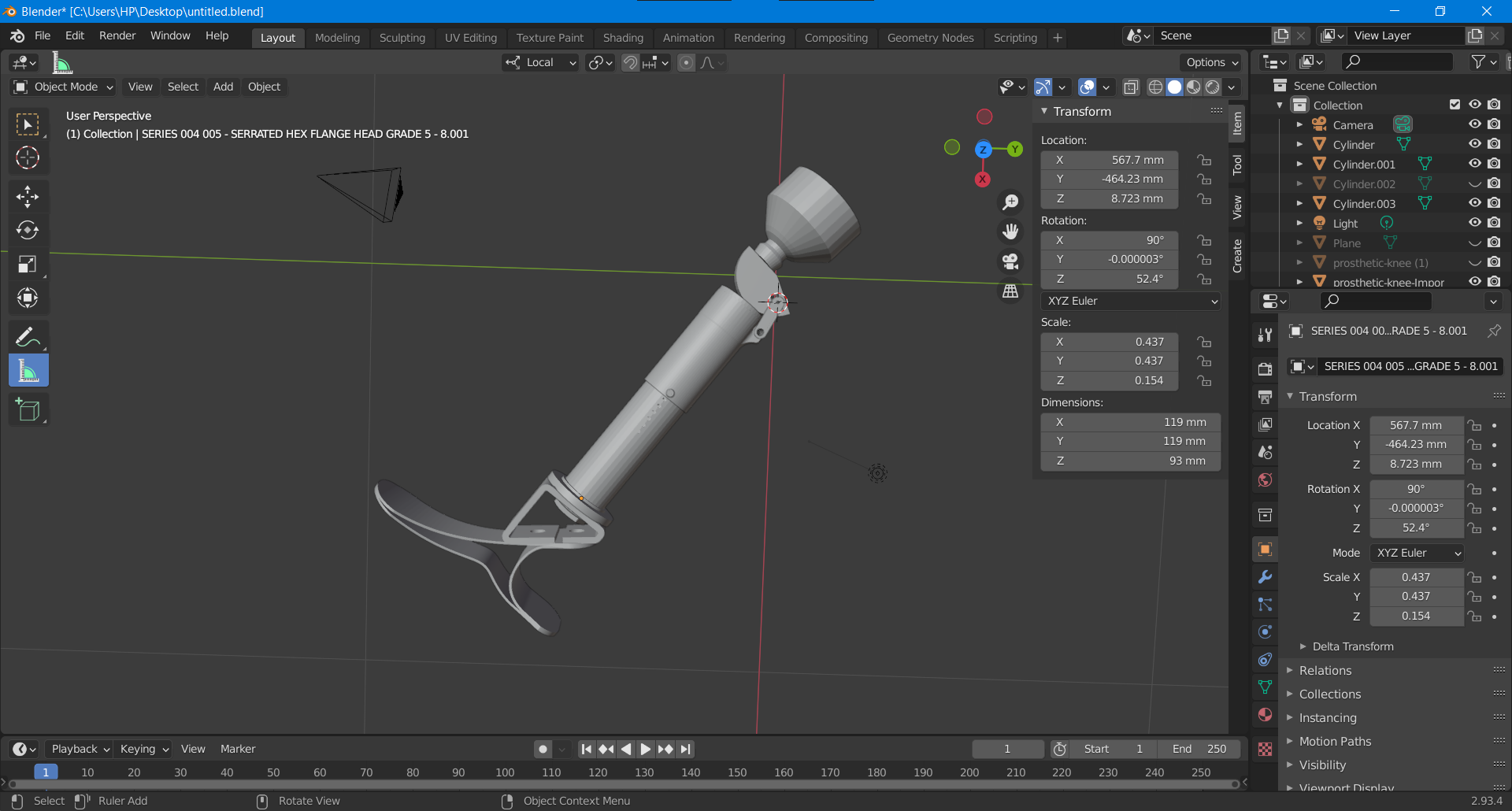}
    \caption{manfucturing process of proshetic leg design after design on solidworks.}
\end{figure}

\begin{figure}[h]
    \centering
    \includegraphics[width=0.4\textwidth]{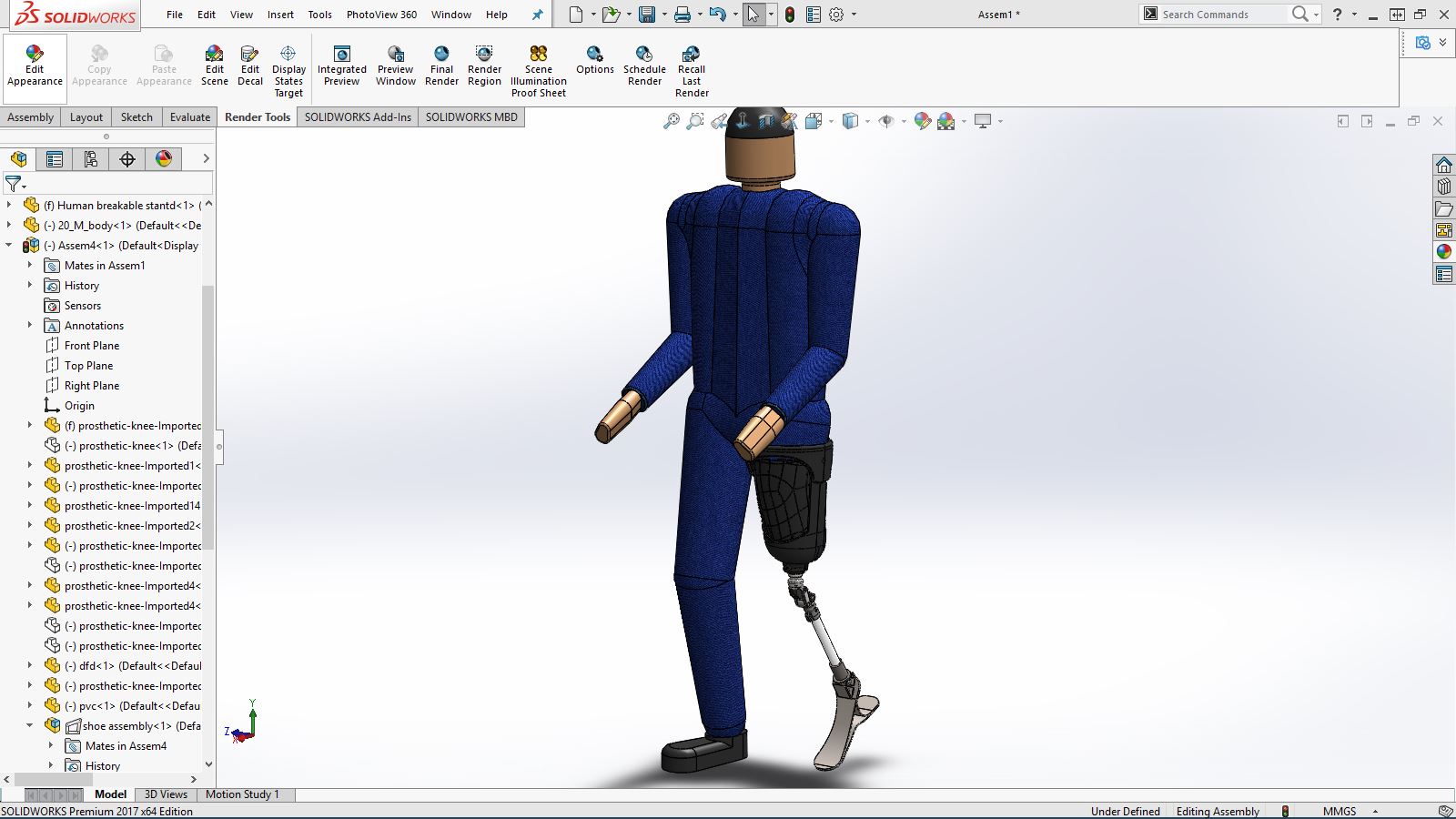}
    \caption{Prosthetic leg controlled by surface EMG. The surface electrodes placed on the user's residual limb capture the electrical signals generated by the muscles.}
\end{figure}

\subsection{Stress Analysis of Presthestic limbs}
The Von-Mises Stress analysis diagram, which is employed to simulate the maximum load that a prosthetic leg can endure. This analysis provides crucial insights into the leg's structural integrity and its ability to withstand external forces. As depicted in the diagram, the calculated maximum load capacity for the leg is approximately 3000 newtons. This information is significant in the field of prosthetics as it aids in designing and developing prosthetic limbs that can safely support the expected loads and stresses experienced during daily activities. By considering the Von-Mises Stress analysis, engineers and researchers can ensure that prosthetic legs are optimized for strength and durability, thereby enhancing their functionality and improving user confidence and safety.
\begin{figure}[h]
    \centering
    \includegraphics[width=0.4\textwidth]{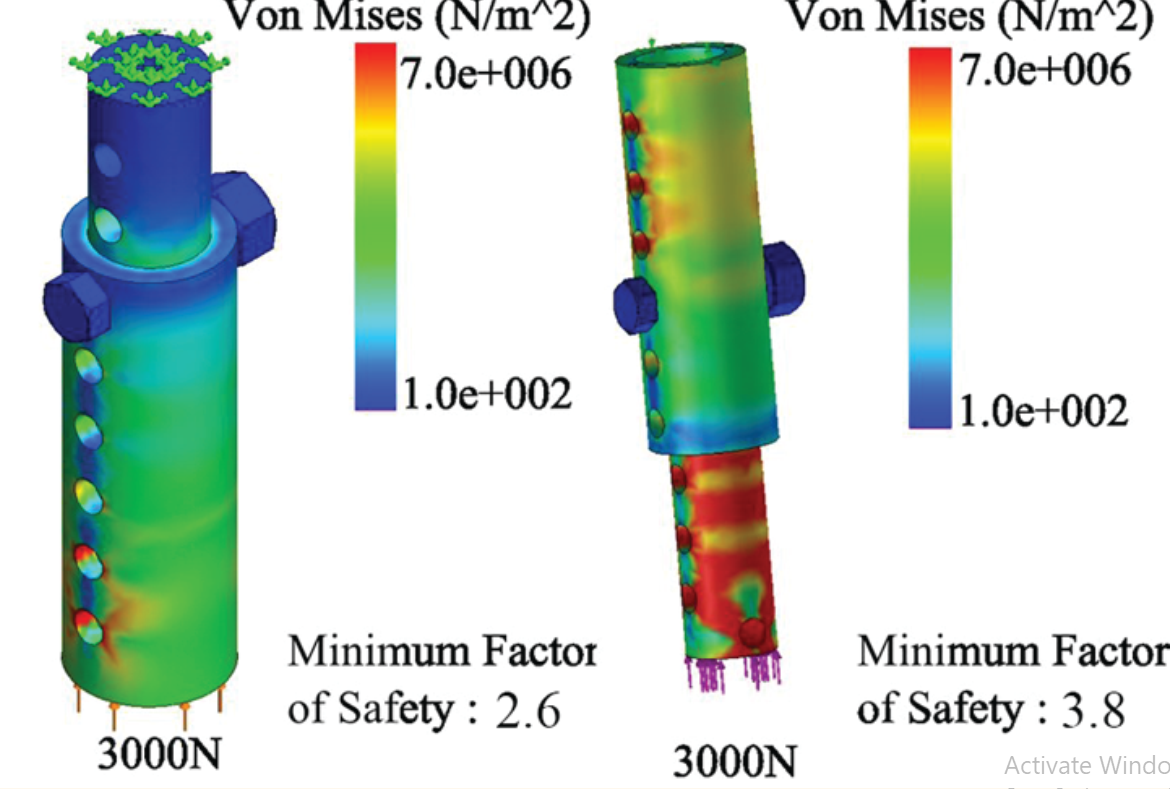}
    \caption{Von-Mises Stress analysis diagram, simulate the maximum load
    which the leg can effort. As shown in figure, the Max load the leg can bear
    approximately 3000 newtons.}
\end{figure}

\begin{figure}[h]
    \centering
    \includegraphics[width=0.45\textwidth]{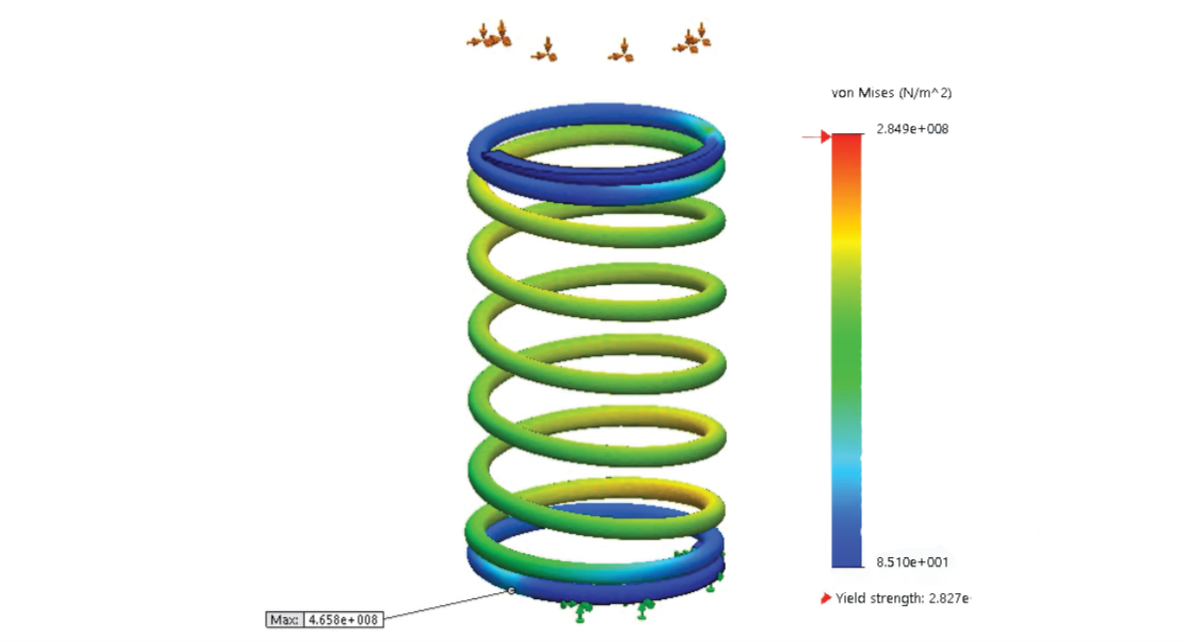}
    \caption{Von-Mises Stress analysis diagram, simulate the maximum load which the leg can effort. As shown in figure, the Max load the leg can bear approximately 3000 newtons.}
\end{figure}

\subsection{Detecting Staris using Machine learning}
Using canny detection algorithm to detect edge, and it analyzed data based on Noise Reduction and Edge detection is sensitive to image noise, the first step is to
eliminate the noise with a 5x5 or 3x3 Gaussian filter. Then, finding the Image’s Intensity Gradient. The smoothed picture is then filtered in both the horizontal
and vertical directions with a Sobel kernel to obtain the first derivative in both the horizontal (Gy) and vertical (Gx) directions.

\begin{equation}
    \text{EdgeGradient}(G) = \sqrt{G^2_x + G^2_y}
\end{equation}

\begin{figure}[h]
    \centering
    \includegraphics[width=0.45\textwidth]{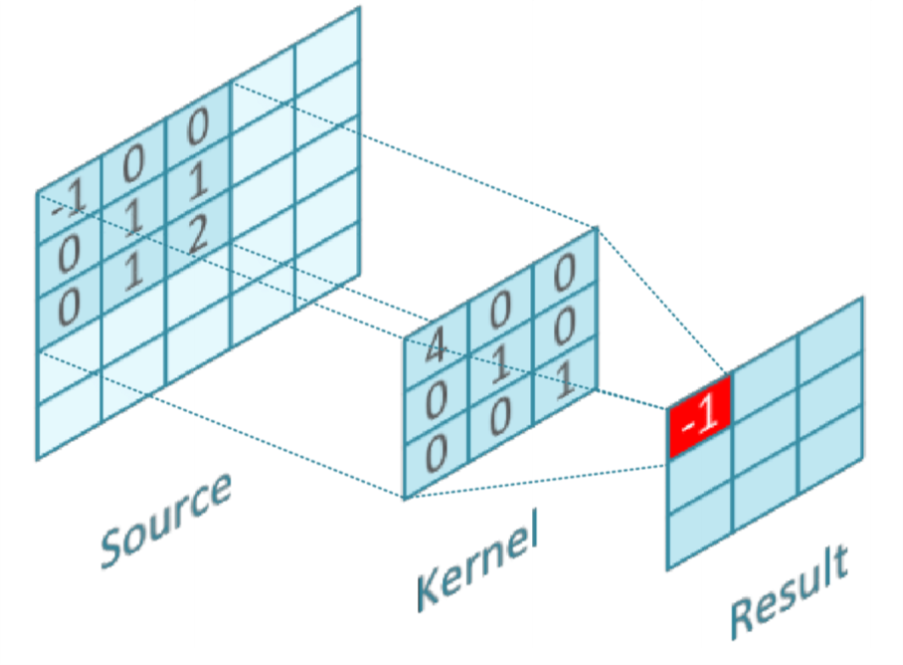}
    \caption{A convolution of a 5x5 image with a 3x3 kernel.}
\end{figure}

\begin{figure}[h]
    \centering
    \includegraphics[width=0.45\textwidth]{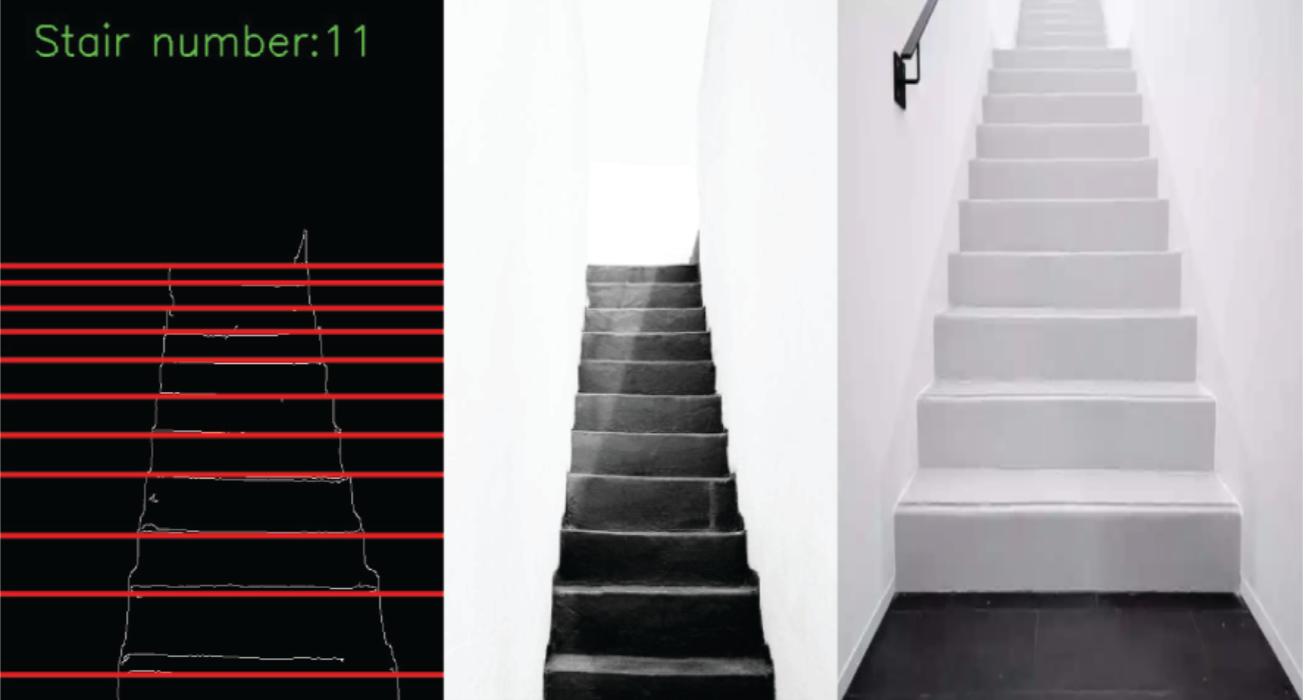}
    \caption{Detecting stairs using machine learning and computer vision is critical for amputees. To detect any stairs, use gaussian blur and the Hough line transform on pictures. We used it to create a prosthetic limb that can move up and down stairs, allowing amputees to function more comfortably with it. It calculates the degree of movement by measuring the height of the stairs.}
\end{figure}

\subsection{Piezoelectric and Voltage generator}
For Powering the prosthetic leg we used Piezoelectric to generate required voltage with each movement,  we stored the overcome voltage in capacitor and battery cells
We collected average 1.48 volts with for each  .45 Kg. Each Trail average consist of 6 trails.
The utilization of piezoelectric devices for battery recharging has garnered significant attention. Extensive testing is imperative to determine their viability as the primary voltage generator. The accompanying graph showcases the relationship between the load and voltage, revealing that as the load increases, the voltage also rises. This observation establishes the adequacy of the device's output as a main voltage generator. Such findings contribute valuable insights into the potential application of piezoelectric devices in sustainable energy harvesting and battery charging systems.

\begin{figure}[h]
    \centering
    \includegraphics[width=0.45\textwidth]{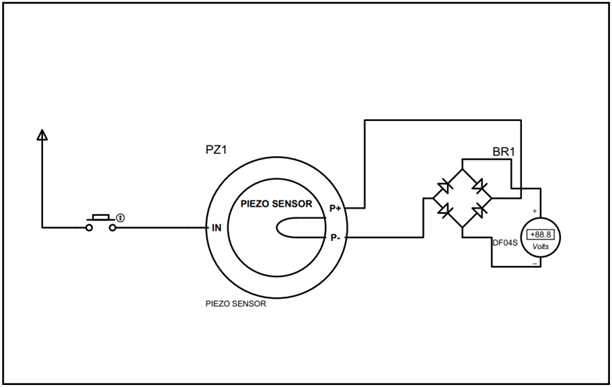}
    \caption{schematic diagrma of piezoelectric and convert from AC to DC.}
\end{figure}

\begin{figure}[h]
    \centering
    \includegraphics[width=0.45\textwidth]{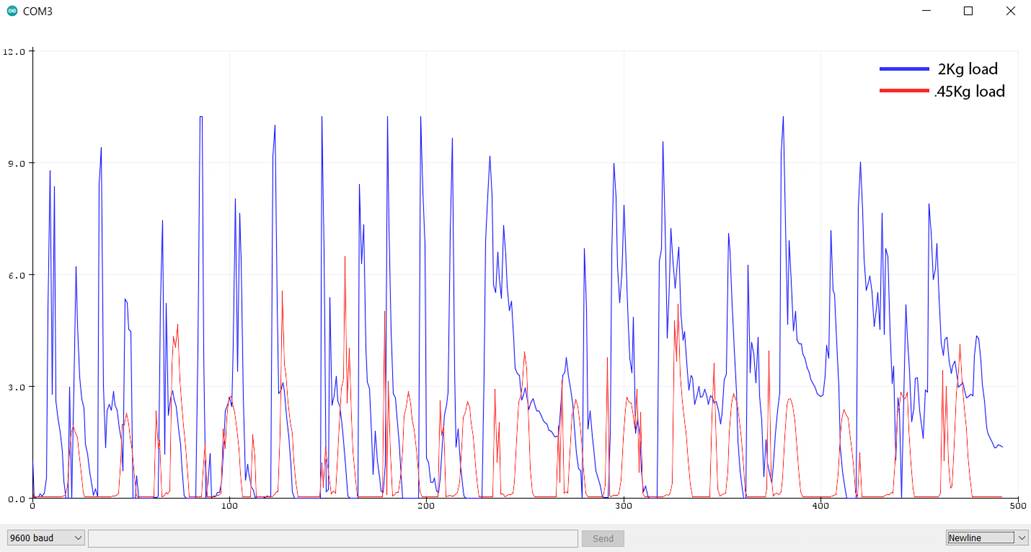}
    \caption{Using piezoelectric devices to recharge batteries. It is necessary to test numerous times to see if it is suitable to be main voltage generator. As demonstrated in the graph, as the load increases, so does the voltage. The output was sufficient for being main voltage generator.}
\end{figure}

\subsection{Early diagnosis of dystonia}
Runner’s dystonia (RD) is a task-specific focal dystonia of
the lower limbs that occurs when running to diagnostic
of dystonia. In this retrospective case series, we present
surface electromyography (EMG) and joint kinematic data
from thirteen patients who underwent instrumented gait analysis (IGA) at the Functional and Biomechanics Laboratory
at the National Institutes of Health[1]. Four cases of RD
are described in greater detail to demonstrate the potential
utility of EMG with kinematic studies to identify dystonia muscle groups in RD. Lateral heel whip, a proposed
novel presentation of lower-limb dystonia, is also described.
Surface EMG is showing continuous activity in the left hamstrings (b) and early activity in the left tibialis anterior during
running (a). Showing how left leg delayed in activation of
motor neuron with respect to other leg
(A) in the time domain and in (B) in the frequency domain.
The signal was recorded from the vastus lateralis (VL) muscle
during Whole body vibration (WBV) at 30 Hz. While the
sEMG signal in the time domain does not highlight any
specific characteristics to WBV. The sEMG signal in the frequency domain clearly shows excessive spikes at the vibration
frequency and at a few multiple harmonics and that means
how dystonia is diagnostic obviously. At the same time, and
also for preliminary purposes, the electrophysiological signal
was recorded from the patella during WBV. Such a signal
obtained during WBV at 30 Hz is shown in Figure 9. In the
time domain, the patella signal resembles a sinusoidal wave
at 30 Hz. In frequency domain, excessive spikes are observed
at the vibration frequency and to a lower extent at its multiple
harmonics. No myoelectrical activity is shown for all the other
frequencies.
\begin{figure}[h]
    \centering
    \includegraphics[width=0.45\textwidth]{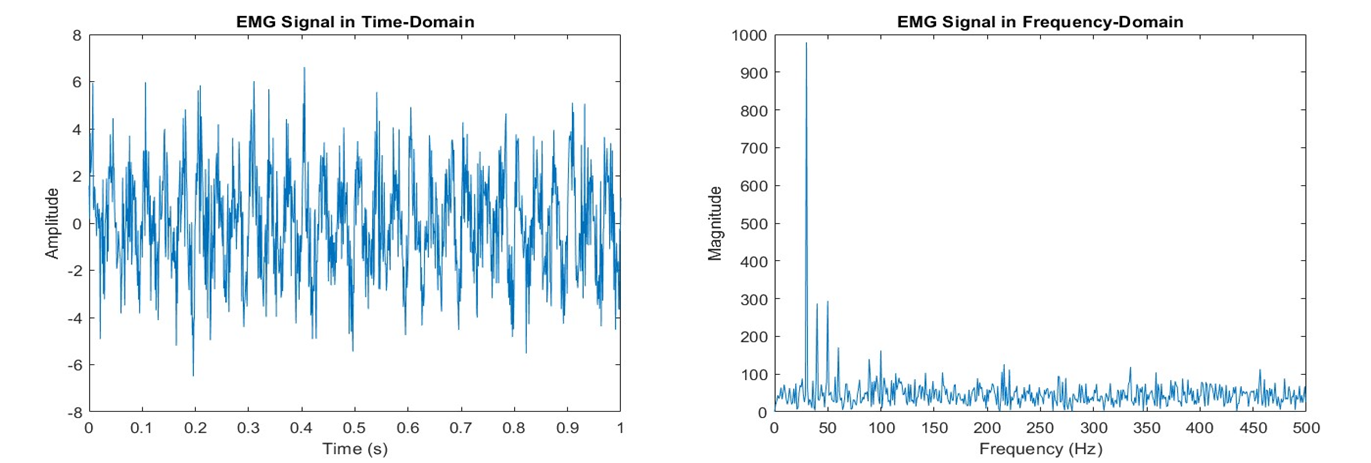}
    \caption{Early diagnosis of Dystonia using sEMG and Fast Fourier Transform (FFT).}
\end{figure}
A surface electromyography (sEMG) spectrum of the Vastus
Lateralis during whole-body vibration at 30 Hz. sEMG signals
were processed using the no-filter method (black solid line),
linear interpolation (grey solid line), band-stop filter (grey
dotted line) and band-pass filter (black dashed line). As shown
in Figure 10, The crucial role of filter especially High Pass
filtration (HPF) and Low Pass filtration (LPF).

\section{Real testing on subjects}

\begin{figure}[htbp]
    \caption{Representing The number of subjects that are tested even public or observed data.}
    \centering
    \includegraphics[width=0.5\textwidth]{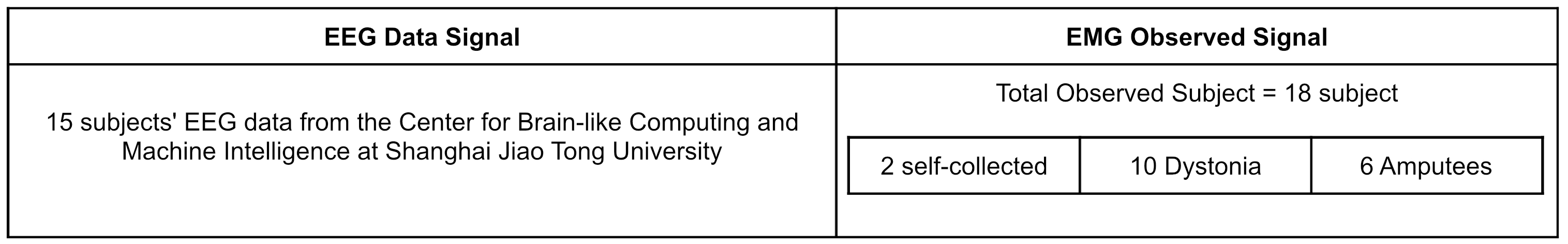}
  \end{figure}

\begin{figure}[h]
    \centering
    \includegraphics[width=0.45\textwidth]{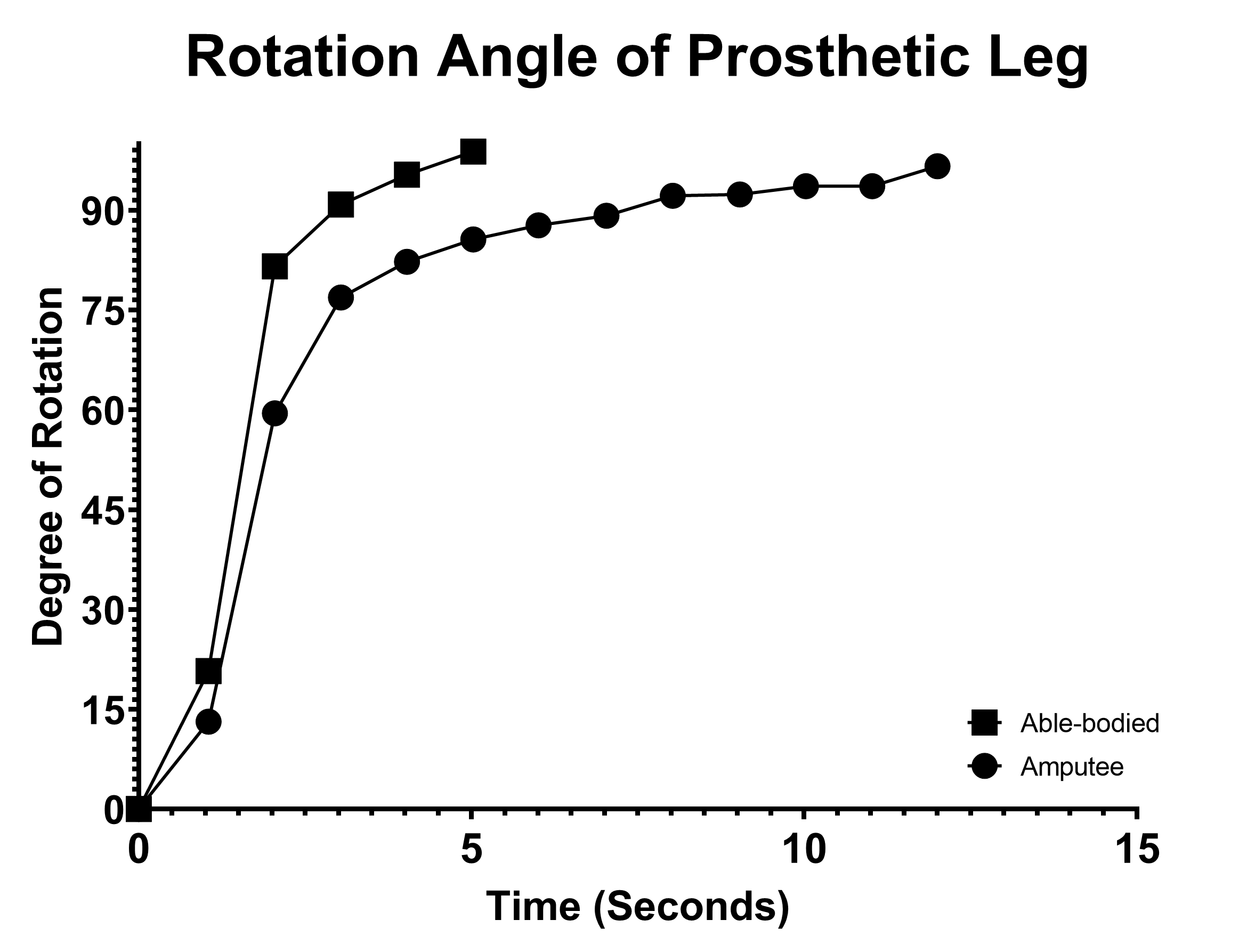}
    \caption{Comparison of angular achievement in prosthetic limb rotation between able-bodied and amputee individuals using sEMG control.}
\end{figure}
Based on the previous discussion, it can be concluded that digital filters overtake analog filters in EMG filtration in terms of flexibility, adaptability, and stability. The utilization of digital filters led to an enhancement of the signal processing stage by approximately 200
The RMSE value obtained from the analog filter was 0.0686, while the digital filter was 0.0327, indicating a significant improvement in signal accuracy. These findings suggest that digital filters are highly effective for EMG filtration and can provide substantial improvements in signal processing performance.

\subsection{Equations}
\begin{equation}
    y(n) = \sum_{i=0}^{M} b_i x(n-i) - \sum_{j=1}^{N} a_j y(n-j)
\end{equation}

An infinite impulse response (IIR) filter is a type of digital filter where the output of the filter depends not only on current and past input samples but also on past output samples.
\begin{figure}[h]
    \centering
    \includegraphics[width=0.45\textwidth]{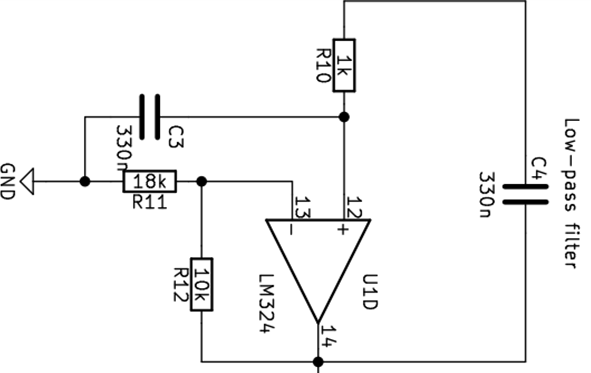}
    \caption{Analog filter schematic diagram.}
\end{figure}
In digital signal processing, the filter function is expressed mathematically, while analog filters are designed using schematic.

\section{Discussion and conclusion}
Digital filters outperformed analog filters for filtering EMG signals, resulting in significant improvements in signal processing performance. Also, VGG16-V1 model was more efficient and accurate than VGG16-V2. These advancements enhanced the use of sEMG signals for controlling prosthetic legs and early diagnosis of dystonia.
\section{Future Plans}
\begin{itemize}
    \item Increase the dataset size and diversity to enhance accuracy and generalization.
    \item User Experience optimization and usability of the sEMG system based on user feedback.
    \item Implement Real-time Feedback by Integrating sensory feedback mechanisms for enhanced control and user experience.
    \item Utilizing more advancing filtration method of digital filtering like wavelet-based or adaptive filters, for more filtering improvements.
\end{itemize}

\end{document}